\documentclass[11pt]{article}
\usepackage[utf8]{inputenc}
\usepackage[T1]{fontenc}
\pdfoutput = 1

\usepackage{amsmath,amssymb,amsthm}
\usepackage{graphicx,subfig}
\usepackage{changepage}
\usepackage{cite,url}
\usepackage{mathtools,xparse,MnSymbol,slashed,mathrsfs}
\usepackage{multirow}
\usepackage{stackrel}
\allowdisplaybreaks

\usepackage{bbm}
\usepackage{bbold}
\usepackage{caption}
\usepackage{cite}
\usepackage{color}
    \definecolor{darkgreen}{rgb}{0,0.5,0}
    \definecolor{darkred}{rgb}{0.5,0,0}
    \definecolor{darkblue}{rgb}{0,0,0.6}
    \definecolor{purple}{rgb}{0.4,.2,0.7}
\usepackage{float}
\usepackage[hyperfootnotes = true, colorlinks = true, linkcolor = darkblue, citecolor = purple]{hyperref}

\numberwithin{equation}{section}

\usepackage[margin = 2.2cm]{geometry}
\usepackage[ragged]{footmisc}
\DeclareMathOperator{\Tr}{Tr}

\def\bra#1{\langle #1 |}
\def\ket#1{| #1 \rangle}

\def\brac#1{\llangle #1 \|}
\def\ketc#1{\| #1 \rrangle}
\def\innerc#1#2{\llangle #1 \| #2 \rrangle}

\begin{document}

\begin{titlepage}
\thispagestyle{empty}

\begin{flushright}
\end{flushright}

\vspace{1.2cm}
\begin{center}

\noindent{\bf \LARGE Complementarity for a Dynamical Black Hole}

\vspace{0.6cm}

{\bf \large Benjamin Concepcion,$^{a,b}$ Yasunori Nomura,$^{a,b,c,d}$ Kyle Ritchie,$^{a,b}$ \\ and Samuel Weiss$^{a,b}$}
\vspace{0.3cm}\\

{\it $^a$ Berkeley Center for Theoretical Physics, Department of Physics, \\
University of California, Berkeley, CA 94720, USA}\\

{\it $^b$ Theoretical Physics Group, Lawrence Berkeley National Laboratory, \\ Berkeley, CA 94720, USA}\\

{\it $^c$ RIKEN Interdisciplinary Theoretical and Mathematical Sciences Program (iTHEMS), \\
Wako, Saitama 351-0198, Japan}\\

{\it $^d$ Kavli Institute for the Physics and Mathematics of the Universe (WPI), \\
UTIAS, The University of Tokyo, Kashiwa, Chiba 277-8583, Japan}\\

\vspace{0.3cm}
\end{center}

\begin{abstract}
Black hole complementarity posits that the interior of a black hole is not independent from its Hawking radiation.
This leads to an apparent violation of causality:\ the interior can be acausally affected by operators acting solely on the radiation.
We argue that this perspective is misleading and that the black hole interior must be viewed as existing in the causal past of the Hawking radiation, despite the fact that they are spacelike separated in the semiclassical description.
Consequently, no operation on the Hawking radiation---no matter how complex---can affect the experience of an infalling observer.
The black hole interior and the radiation only appear spacelike separated in the semiclassical description because an infalling observer's ability to access complex information is limited; the chaotic dynamics on the horizon, as viewed from the exterior, then converts any effect caused by such an observer to information in the Hawking radiation which cannot be accessed at the semiclassical level. 
We arrive at the picture described above by considering a unitary exterior description in which the flow of information is strictly causal, which we extend to apply throughout the entire history of black hole evolution, including its formation.
This description uses the stretched event horizon as an inner edge of spacetime, on which the information inside is holographically encoded.
We argue that the global spacetime picture arises from coarse-graining over black hole microstates, and discuss its relationship with the exterior description.
\end{abstract}

\end{titlepage}

\tableofcontents
\newpage

\section{Introduction}
\label{sec:intro}

Studying the quantum mechanics of a black hole has revealed surprising facts about the nature of spacetime~\cite{Bekenstein:1973ur,Bekenstein:1974ax,Hawking:1974sw}.
Of particular interest is that the black hole interior and Hawking radiation are not independent quantum degrees of freedom.
This idea, called complementarity~\cite{Susskind:1993if}, has been tested in various simplified setups~\cite{Penington:2019npb,Almheiri:2019psf,Almheiri:2019hni,Penington:2019kki,Almheiri:2019qdq}, using methods~\cite{Ryu:2006bv,Hubeny:2007xt,Faulkner:2013ana,Engelhardt:2014gca} developed by analyzing the holographic nature of quantum gravity~\cite{tHooft:1993dmi,Susskind:1994vu}, in particular the AdS/CFT correspondence~\cite{Maldacena:1997re}.

This lack of independence allows us to construct operators representing the interior which act solely on the Hawking radiation, as is indeed implied by entanglement wedge reconstruction~\cite{Czech:2012bh,Wall:2012uf,Headrick:2014cta,Jafferis:2015del,Dong:2016eik,Faulkner:2017vdd,Cotler:2017erl,Akers:2020pmf} with entanglement islands~\cite{Penington:2019npb,Almheiri:2019psf,Almheiri:2019hni}.
Since the radiation and the interior are spacelike separated, an exterior observer's ability to affect interior degrees of freedom leads to an apparent violation of causality.
This issue was analyzed by Kim, Tang, and Preskill~\cite{Kim:2020cds}, who studied a partially evaporating black hole and proposed that the spacelike nature of the interior and radiation in the semiclassical description is protected because the operators representing the interior effectively commute with any operators that do not have computational complexity exponential in the entropy of the black hole.
They, however, did not address how to interpret operators acting on the radiation after the black hole is fully evaporated, which would still be able to affect the interior acausally.
Bousso and Penington studied a process involving such operators and concluded that it leads to a breakdown of the semiclassical spacetime~\cite{Bousso:2023sya}.

In this paper, we study how complementarity is manifested without the issue of causality described above.
We claim that to fully understand this issue, we need to properly analyze how a dynamical black hole behaves under time evolution.
For this purpose, we adopt an exterior description~\cite{Nomura:2018kia,Nomura:2019qps,Nomura:2019dlz,Langhoff:2020jqa,Nomura:2020ska,Murdia:2022giv} in which an effective theory describing a portion of the black hole interior is constructed from the horizon (and Hawking radiation) degrees of freedom \textit{at each exterior time}.
This portion of the interior is obtained by evolving the modes on the corresponding exterior time slice by the \textit{infalling} Hamiltonian, not by exterior time evolution, and therefore must be viewed as associated with the exterior time at which the effective theory is erected.
A sketch of this construction is given in Fig.~\ref{fig:eff-int-region}.
In this description, the black hole interior is a transient phenomenon obtained by reshuffling the relevant degrees of freedom while an infalling object is being scrambled into the black hole.

Using this construction, we show the following:
\begin{itemize}
\item
The black hole interior must be viewed as existing in the causal past of the corresponding Hawking radiation despite the fact that they are spacelike separated in the semiclassical picture.
In the fundamental description, the relationship between operators corresponding to the interior and their reconstruction in the Hawking radiation is timelike.
\item
The relationship between these operators becomes effectively spacelike in the semiclassical description because of limitations on an infalling observers' ability to access complex information.
With this limitation, the chaotic dynamics on the horizon, as viewed from the exterior, convert any effect caused by an interior observer to information in the Hawking radiation which cannot be accessed at the semiclassical level.
\item
An operation on the Hawking radiation cannot affect the experience of an observer in the interior no matter how complex it is, providing a solution to the causality issue different from that in~\cite{Kim:2020cds}.
Executing such an operation can only hope to ``rewrite the record'' to make it appear that the past was different.
We show that the same applies also to black hole mining~\cite{Unruh:1982ic,Brown:2012un}, so that the semiclassical spacetime is robust under information extraction.
\end{itemize}
Note that this provides a simple solution to the problem of possible information cloning in the interior and the radiation.
The no-cloning theorem~\cite{Wootters:1982zz} is not violated simply because an infalling object and its information in Hawking radiation are the same timelike-related degrees of freedom.

We will formulate a unitary exterior description in which the flow of information is strictly causal throughout the black hole's formation and evaporation.
This is not trivial, especially during the black hole formation era.
We will describe the challenge and propose a solution which applies bulk holography to the stretched \textit{event} horizon.
In this description, the stretched event horizon provides the inner ``edge'' of the spacetime, on which the information about the interior is holographically encoded.

The picture presented here implies that a description based on the global spacetime is necessarily coarse-grained.
We will discuss the relation between the exterior description and calculations employing the global spacetime description, such as replica wormhole calculation~\cite{Penington:2019kki,Almheiri:2019qdq}, entanglement islands~\cite{Penington:2019npb,Almheiri:2019psf,Almheiri:2019hni}, and $\alpha$-states arising from baby universes~\cite{Coleman:1988cy,Giddings:1988cx,Giddings:1988wv} created by the evaporation of a black hole~\cite{Marolf:2020rpm,Polchinski:1994zs}.

The organization of this paper is as follows.
After laying out the exterior description in Section~\ref{sec:exterior}, we discuss in Section~\ref{sec:emergence} how the interior of the black hole and the global spacetime of general relativity emerge from the exterior description.
In Section~\ref{sec:puzzle-sol}, we argue that the global spacetime description cannot be used to define unitary quantum mechanics consistent with interior reconstruction, and show how the framework in Section~\ref{sec:emergence} addresses this issue.
In particular, we see that the black hole interior must be viewed as existing in the past of the Hawking radiation containing its information, despite the fact that they are spacelike separated in the semiclassical description.
We discuss how the timelike relation can be regarded as spacelike at the semiclassical level because of the limitation of an infaller's ability and the chaotic dynamics of the horizon degrees of freedom.

In Section~\ref{sec:causal}, we extend the framework of Section~\ref{sec:emergence} to the entire history of black hole evolution, including its formation era.
This gives a description in which the flow of information is causally trivial throughout the history.
Our construction employs bulk holography associated with the stretched event horizon.
In Section~\ref{sec:global}, we discuss the relation of our picture with other works that use the global spacetime description.
Section~\ref{sec:concl} gives conclusions.

Throughout the paper, we adopt natural units $\hbar = c = k = 1$.

\section{Exterior Description}
\label{sec:exterior}

Black holes form and then evaporate, and a distant observer expects the evolution to be unitary in agreement with quantum mechanics \cite{tHooft:1984kcu,tHooft:1990fkf,Susskind:1993if,Page:1993wv}.
Unitarity is further supported by the AdS/CFT correspondence\cite{Maldacena:1997re}.
To preserve unitarity, we must follow \textit{all} the relevant degrees of freedom, including the ``intrinsically quantum gravitational'' (or ``Planckian'') degrees of freedom associated with the stretched horizon~\cite{Susskind:1993if}.
For a static black hole, this horizon is located at a proper distance of order the cutoff/string length $l_{\rm s}$ away from the mathematical horizon.
The spacetime region relevant for the description is the exterior of the stretched horizon.

\begin{figure}[t]
\centering
  \includegraphics[height=0.37\textheight]{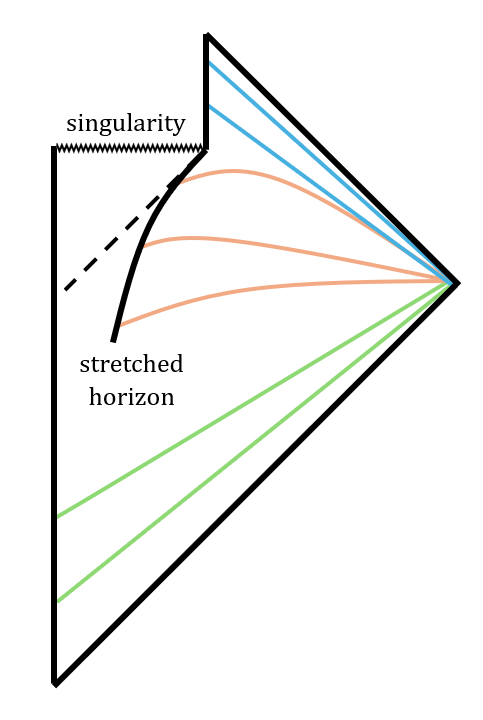}
\vspace{-2mm}
\caption{
 In the exterior description, equal-time slices foliate the spacetime region outside the ``inner boundary'' of space, which for a (quasi-)static black hole corresponds to the stretched horizon.
}
\label{fig:exterior-slice}
\end{figure}

A simple way to implement the exterior description is to adopt the canonical formalism of quantum mechanics.
The relevant time slices are Schwarzschild-like time slices, i.e.\ slices that foliate the spacetime region outside the horizon; see Fig.~\ref{fig:exterior-slice}.
At times when the black hole exists, these time slices cover the region outside the black hole.
In the context of holography, these slices can be obtained, for example, by continuously pulling the boundary into the bulk using the method in~\cite{Nomura:2018kji,Murdia:2020iac}.

A feature of the exterior description is that if the state before the formation of the black hole is pure, then the states with the black hole present and those after it evaporates are also pure.
The relevant degrees of freedom in each period are
\begin{itemize}
\item
{\bf Before the black hole formation:}\
low energy degrees of freedom whose dynamics can be described by a semiclassical theory.
In particular, they include matter that collapses gravitationally to form the black hole.
\item
{\bf While the black hole is present:}\
the degrees of freedom outside the horizon.
At each time, we can divide these degrees of freedom into three classes (see~\cite{Murdia:2022giv} for more detail):
\begin{itemize}
\item
{\it Horizon modes:} modes on the stretched horizon, which may be viewed as ``located'' in the region $r \leq r_{\rm s}$, where $r_{\rm s}$ represents the location of the stretched horizon in the Schwarzschild radial coordinate $r$ (although the classical spacetime picture is not valid in this region).%
\footnote{
 Without an explicit ultraviolet theory, it is scheme dependent if we view these degrees of freedom as additional degrees of freedom beyond the low energy quantum fields or their cutoff-scale excitations near the stretched horizon.
 In any event, the microscopic dynamics of these degrees of freedom cannot be described by low energy semiclassical theory.
}
\item
{\it Zone modes:} modes localized in the region $r_{\rm s} < r \leq r_{\rm z}$, where $r_{\rm z}$ is the location of the outer edge of the near black-hole region called the zone (i.e.\ the region inside the barrier of the effective gravitational potential generated by the black hole).
\item
{\it Far modes:} modes located in the asymptotic region, $r > r_{\rm z}$.
\end{itemize}
This decomposition can be performed at the gauge-fixed level with positive semi-definite inner products between the states, up to the intrinsic ambiguity of separation between horizon and zone modes
(we assume that the usual issue of factorizing modes located in different regions in quantum field theory is appropriately taken care of).
The dynamics of the horizon modes cannot be described by a low energy semiclassical theory, while those of the zone and far modes can.
\item
{\bf After the black hole evaporation:}\
low energy degrees of freedom whose dynamics can be described by the semiclassical theory.%
\footnote{
 This does not mean that the state of these degrees of freedom is fully specified by operators within the low energy semiclassical theory; see Section~\ref{sec:puzzle-sol}.
}
In particular, they include Hawking radiation emitted from the black hole.
\end{itemize}

As has long been argued~\cite{tHooft:1984kcu,tHooft:1990fkf,Susskind:1993if,Page:1993wv}, the exterior description should be unitary, since the stretched horizon behaves as a regular material surface as far as the flow of quantum information is concerned.
On the other hand, general relativity implies the existence of an interior.
While there are proposals otherwise, such as fuzzballs~\cite{Lunin:2001jy,Lunin:2002iz,Mathur:2003hj} and firewalls~\cite{Almheiri:2012rt,Almheiri:2013hfa,Marolf:2013dba}, we take the view that nothing special happens at the horizon for an infalling observer.
The question then is how we can understand the emergence of a smooth interior for the infaller in the exterior description.

\section{Emergence of Global Spacetime}
\label{sec:emergence}

In~\cite{Nomura:2018kia,Nomura:2019qps,Nomura:2019dlz,Langhoff:2020jqa,Nomura:2020ska,Murdia:2022giv}, it was shown that an effective theory of a smooth interior can be constructed from the exterior degrees of freedom.
In particular, the smooth interior emerges as a secondary concept associated with the modes in the zone describing an object falling into the black hole.
Coherent excitations of the rest of the zone and horizon modes (as well as far modes for an old black hole) play the role of the modes in the second exterior of a two-sided geometry, which is emergent in the effective theory.
As emphasized in~\cite{Nomura:2018kia,Nomura:2019qps,Nomura:2019dlz,Langhoff:2020jqa,Nomura:2020ska,Murdia:2022giv}, it is important to apply this construction to a state \textit{at a fixed time $t$} in the exterior description.
This provides \textit{a portion of} the black hole interior, which can be used to describe the experience of the infalling object at the semiclassical level.
To cover the entire black hole interior, we need multiple effective theories erected at different times.

In Section~\ref{subsec:eff-int}, we review this construction.
We will focus on the main idea, referring the readers to~\cite{Murdia:2022giv} for more details.
In Section~\ref{subsec:comp-cutoff}, we discuss properties of the construction relevant for our later discussions, especially focusing on how the picture of global spacetime may emerge.

\subsection{Effective theory of the interior}
\label{subsec:eff-int}

For concreteness, we focus on a collapse-formed black hole with negligible spin and charge in 4-dimensional asymptotically flat spacetime.
We assume that the black hole has a mass $M$ at some time $t$, where $t$ is the time in the exterior description.%
\footnote{
 More precisely, we focus on a branch of the state at $t$ in which the black hole has mass $M$ and is located at a fixed position, up to irreducible quantum uncertainties of order the Hawking temperature $T_{\rm H}$ and the proper length $l_{\rm s}$, respectively~\cite{Nomura:2018kia}.
 Our construction of the effective theory of the interior applies to each such branch separately.
}
The horizon, zone, and far modes are defined based on their locations \textit{at this time}, according to the criterion given in Section~\ref{sec:exterior}.
A \textit{vacuum} microstate of the semiclassical black hole at time $t$, i.e.\ a state in which excitations described by the semiclassical theory are absent, can then be written as follows.

We first select a subset of the zone modes that will be relevant for describing a falling object which is assumed to lie in the zone at time $t$.
These are called \textit{hard modes}, which we collectively label as $\alpha$.
The semiclassical vacuum microstate, together with Hawking radiation emitted earlier, can then be written as%
\footnote{
 Here we ignore the effect of backreaction of Hawking emission on the occupation numbers of hard modes, which has some implication for the region that can be directly reconstructed from the state $\ket{\Psi(M)}$ without resorting to exterior time evolution; for details, see~\cite{Murdia:2022giv}.
}
\begin{equation}
  \ket{\Psi(M)} = \sum_n \sum_{i_n = 1}^{e^{S_{\rm bh}(M-E_n)}} \sum_a c_{n i_n a} \ket{\{ n_\alpha \}} \ket{\psi^{(n)}_{i_n}} \ket{\phi_a},
\label{eq:BH-state}
\end{equation}
where $\ket{\{ n_\alpha \}}$ represent states of the hard modes, specified by the occupation number $n_\alpha$ for each mode $\alpha$, while $\ket{\psi^{(n)}_{i_n}}$ represent orthonormal states of the horizon and zone modes that are not included in the hard modes, which are called \textit{soft modes} (any basis in the $i_n$ space can be taken for each $n$).
Finally, $\ket{\phi_a}$ are the states of the far modes, which generally include Hawking radiation emitted earlier.
Note that since the total energy of the black hole system, consisting of the hard and soft modes, is $M$, the number of basis states $\ket{\psi^{(n)}_{i_n}}$ appearing with $\ket{\{ n_\alpha \}}$ in~\eqref{eq:BH-state} is given by $S_{\rm bh}(M-E_n)$.
Here, $E_n$ is the energy carried by $\ket{\{ n_\alpha \}}$, $S_{\rm bh}(E) = 4\pi E^2 l_{\rm P}^2$ is the Bekenstein-Hawking entropy density at energy $E$, and $l_{\rm P}$ is the Planck length, which we take to be not too different from $l_{\rm s}$, assuming that the number of low energy species is not enormous.

\begin{figure}[t]
\centering
  \includegraphics[height=0.3\textheight]{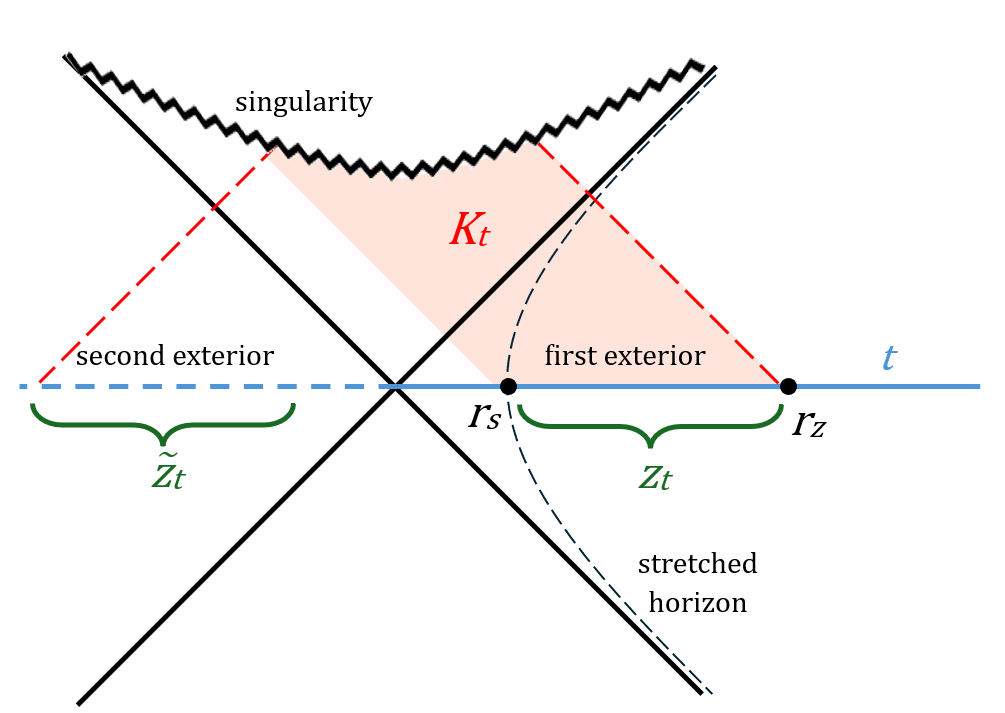}
\vspace{0mm}
\caption{
 The effective theory of the interior erected at time $t$ gives an emergent two-sided black hole geometry.
 The degrees of freedom associated with the zone $\tilde{Z}_t$ of the second exterior arise from soft and far mode states coupled to the states in the zone $Z_t$ of the first exterior in the vacuum microstate.
 The spacetime region described by this effective theory is the region indicated as $K_t$.
}
\label{fig:two-sided}
\end{figure}

Because of fast-scrambling~\cite{Hayden:2007cs,Sekino:2008he} and chaotic~\cite{Maldacena:2015waa} dynamics at the horizon, (the relevant portion of) the coefficients $c_{n i_n a}$ in~\eqref{eq:BH-state} take random values across all low energy species.%
\footnote{
 More precisely, for each $a$, the coefficients $c_{n i_n a}$ are expected to behave as independent and identically distributed complex Gaussian random variables with mean $0$.
}
(The specific values of $c_{n i_n a}$ correspond to a specific black hole microstate.)
This implies that the state of the system can be written in the thermofield double form
\begin{equation}
  \ket{\Psi(M)} = \frac{1}{\sqrt{\sum_m e^{-\frac{E_m}{T_{\rm H}}}}} \sum_n e^{-\frac{E_n}{2T_{\rm H}}} \ket{\{ n_\alpha \}} \ketc{\{ n_\alpha \}},
\label{eq:TFD}
\end{equation}
up to exponential accuracy, where $T_{\rm H} = 1/( \partial S_{\rm bh}(E)/\partial E|_{E=M}) = 1/8\pi M l_{\rm P}^2$ is the Hawking temperature.
The state with the double-ket symbol is defined as the state of soft and far modes which couples to the state $\ket{\{ n_\alpha \}}$ in~\eqref{eq:BH-state} 
\begin{equation}
  \ketc{\{ n_\alpha \}} = \sqrt{\sum_m e^{-\frac{E_m}{T_{\rm H}}}} e^{\frac{E_n}{2T_{\rm H}}}  
  \sum_{i_n = 1}^{e^{S_{\rm bh}(M-E_n)}} \sum_a c_{n i_n a} \ket{\psi^{(n)}_{i_n}} \ket{\phi_a},
\label{eq:ketc}
\end{equation}
where the overall coefficient is determined by the normalization condition, $\innerc{\{ n_\alpha \}}{\{ n_\alpha \}} = 1$.
We see that the double-ket states play the role of the states in the second exterior of the effective two-sided black hole, as depicted in Fig.~\ref{fig:two-sided}.
This allows us to define, along the lines of~\cite{Papadodimas:2012aq,Verlinde:2012cy,Nomura:2012ex,Maldacena:2013xja,Papadodimas:2013jku}, annihilation and creation operators
\begin{alignat}{3}
  & b_\gamma = \sum_n \sqrt{n_\gamma}\, 
    \ket{\{ n_\alpha - \delta_{\alpha\gamma} \}} \bra{\{ n_\alpha \}},
\qquad &
  & b_\gamma^\dagger = \sum_n \sqrt{n_\gamma + 1}\, 
    \ket{\{ n_\alpha + \delta_{\alpha\gamma} \}} \bra{\{ n_\alpha \}},
\label{eq:ann-cre-b}
\\
  & \tilde{b}_\gamma = \sum_n \sqrt{n_\gamma}\, 
    \ketc{\{ n_\alpha - \delta_{\alpha\gamma} \}} \brac{\{ n_\alpha \}},
\qquad &
  & \tilde{b}_\gamma^\dagger = \sum_n \sqrt{n_\gamma + 1}\, 
    \ketc{\{ n_\alpha + \delta_{\alpha\gamma} \}} \brac{\{ n_\alpha \}},
\label{eq:ann-cre-b-tilde}
\end{alignat}
from which we can construct operators describing the interior of the black hole, using only the degrees of freedom on the exterior region, i.e.\ the hard, soft, and far modes defined on the time slice $t$.%
\footnote{
 While the operators in~\eqref{eq:ann-cre-b-tilde} depend sensitively on the microstate of the black hole, i.e.\ the precise values of $c_{n i_n a}$ in~\eqref{eq:BH-state}, we can promote them to apply to a wide range of microstates to the extent that it is sufficient to ensure the validity of the semiclassical description of the interior~\cite{Nomura:2020ska,Papadodimas:2015jra,Hayden:2018khn}.
}

The operators describing semiclassical excitations in the interior region, obtained from Eqs.~\eqref{eq:ann-cre-b} and \eqref{eq:ann-cre-b-tilde}, act on all of the hard, soft, and far modes.
However, if the black hole is young, i.e.\ if the combined system of the hard and soft modes is not maximally entangled with far modes at time $t$, then the interior operators can be constructed without involving far modes~\cite{Nomura:2019dlz,Nomura:2020ska}.
This can be done by projecting the operators acting on all modes onto the Hilbert space of hard and soft modes using the Petz map.
Schematically, when the entanglement between the black hole and far modes is negligible, the vacuum microstate can be written as
\begin{equation}
  \ket{\Psi(M)} \approx \sum_n \sum_{i_n = 1}^{e^{S_{\rm bh}(M-E_n)}}\!\!\! c_{n i_n} \ket{\{ n_\alpha \}} \ket{\psi^{(n)}_{i_n}}.
\label{eq:BH-state-2}
\end{equation}
The double-ket states can then be defined by
\begin{equation}
  \ketc{\{ n_\alpha \}} \approx \sqrt{\sum_m e^{-\frac{E_m}{T_{\rm H}}}} e^{\frac{E_n}{2T_{\rm H}}}  
  \sum_{i_n = 1}^{e^{S_{\rm bh}(M-E_n)}}\!\!\! c_{n i_n} \ket{\psi^{(n)}_{i_n}},
\label{eq:ketc-2}
\end{equation}
instead of~\eqref{eq:ketc}, giving interior operators acting only on hard and soft modes.
For a young black hole, these operators reproduce the same algebra as that obtained using the ``full'' double-ket states in~\eqref{eq:ketc}, up to exponentially suppressed corrections.

We stress that in the case of either a young or an old black hole, interior operators constructed as above \textit{must} involve soft modes, i.e.\ modes near the black hole region.
This implies, in particular, that the construction does not work when a black hole does not exist at time $t$.
This is different from entanglement wedge reconstruction~\cite{Czech:2012bh,Wall:2012uf,Headrick:2014cta,Jafferis:2015del,Dong:2016eik,Faulkner:2017vdd,Cotler:2017erl,Akers:2020pmf}, which allows for reconstructing interior operators only using far modes (early radiation)~\cite{Penington:2019npb,Almheiri:2019psf,Almheiri:2019hni}.
The difference comes from the fact that the construction described here does not use exterior time evolution with respect to $t$ in converting the fundamental modes to infalling modes, while entanglement wedge reconstruction (implicitly) uses it by assuming the knowledge of microscopic time evolution~\cite{Langhoff:2020jqa,Murdia:2022giv}.

For an evaporating black hole, the spacetime region $K_t$ that can be obtained by the construction above is a portion of the interior which corresponds to the intersection of the domain of dependence of the union of $Z_t$ and $\tilde{Z}_t$ with the causal future of $Z_t$~\cite{Murdia:2022giv}:
\begin{equation}
  K_t = D(Z_t \cup \tilde{Z}_t) \cap J^+(Z_t),
\label{eq:K_t}
\end{equation}
where $Z_t$ represents the zone at time $t$, i.e.\ the partial Cauchy surface $r_{\rm s} < r \leq r_{\rm z}$ at $t$, and $\tilde{Z}_t$ is its second exterior mirror in the effective two-sided geometry arising in the effective theory; see Fig.~\ref{fig:two-sided}.
Any operator of the semiclassical theory contained in this region can be constructed using the time evolution operator $\tilde{U}(\tau) = e^{-i \tilde{H} \tau}$ associated with an infalling frame.
Here,
\begin{equation}
  \tilde{H} = \sum_\xi \Omega_\xi a_\xi^\dagger a_\xi + \tilde{H}_{\rm int}\bigl( \{ a_\xi \}, \{ a_\xi^\dagger \} \bigr)
\label{eq:tilde-H}
\end{equation}
is the generator of the infalling time evolution, and $\tau$ is the infalling time; $a_\xi$ and $a_\xi^\dagger$ are the annihilation and creation operators for infalling mode $\xi$, obtained from $b_\gamma$, $b_\gamma^\dagger$, $\tilde{b}_\gamma$, and $\tilde{b}_\gamma^\dagger$ by Bogoliubov transformations, and $\Omega_\xi$ is the frequency of the mode $\xi$.
Note that this time evolution is \textit{different} from the original time evolution, which acts on all the microscopic degrees of freedom.
In fact, the effective theory of the interior obtained in this way is intrinsically semiclassical, possessing intrinsic ambiguities of order $e^{-S_{\rm tot}/2}$, where $S_{\rm tot}$ is the coarse-grained entropy of the combined black hole and radiation system~\cite{Nomura:2020ska}.

The picture presented here implies that a generic state in the Hilbert space of microstates for a collapse-formed, single-sided black hole does not have a firewall.
In~\cite{Marolf:2013dba}, it was argued that since on a black hole background ``arbitrarily many'' configurations of matter can be excited without costing energy as viewed from a distance (by putting them near the horizon), a generic state of the black hole Hilbert space has a firewall.
These configurations, however, correspond precisely to black hole microstates; specific values of $c_{n i_n}$ in~\eqref{eq:BH-state-2}, or $c_{n i_n a}$ in~\eqref{eq:BH-state}, correspond simply to a specific \textit{vacuum} microstate.
For generic values of these coefficients, the present construction works, giving an effective description in which a falling object goes through the horizon smoothly.
In particular, the fate of a semiclassical object in the zone, represented by a set of $b_\gamma^\dagger$ operators \textit{acted on} the vacuum microstate, can be calculated by evolving the system by $\tilde{U}(\tau)$ using the in-in formalism, reproducing the result consistent with general relativity.
It is only for an exponentially atypical state (e.g.\ a state in which one of the coefficients is exponentially larger than the others) that this construction does not work, leading to a firewall at the horizon.%
\footnote{
 This can be phrased such that a generic single-sided state is mapped onto a special (thermofield double) two-sided state in the effective description.
 This occurs because of the state-dependent~\cite{Nomura:2020ska,Papadodimas:2012aq,Verlinde:2012cy,Nomura:2012ex,Maldacena:2013xja,Papadodimas:2013jku,Papadodimas:2015jra,Hayden:2018khn} definition of infalling mode operators.
 For refutations against other arguments~\cite{Mathur:2009hf,Almheiri:2012rt,Almheiri:2013hfa,Bousso:2013ifa} for firewalls, see~\cite{Nomura:2018kia,Nomura:2019qps,Nomura:2019dlz,Langhoff:2020jqa,Nomura:2020ska,Murdia:2022giv} and references therein.
}

\subsection{Global spacetime from a patchwork}
\label{subsec:comp-cutoff}

\begin{figure}[t]
\centering
  \includegraphics[height=0.43\textwidth]{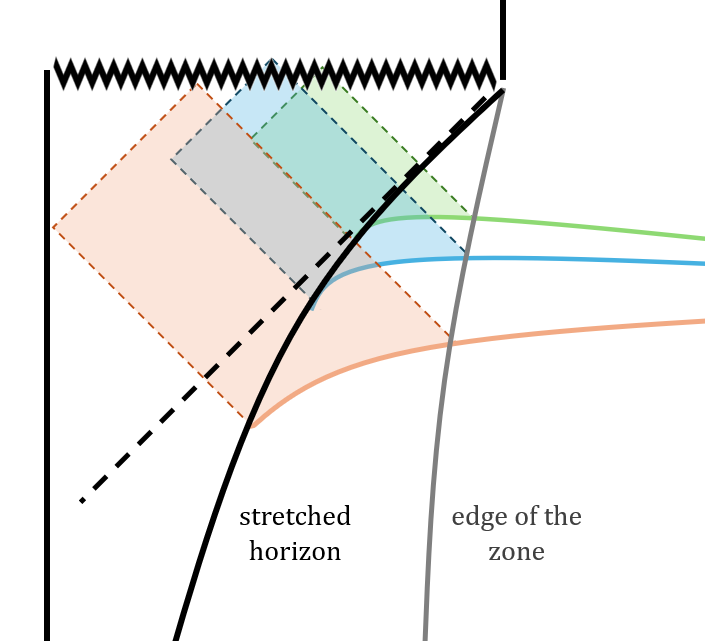}
\vspace{1mm}
\caption{
 The effective theory of the interior erected at an exterior time $t$ describes the region of the black hole interior corresponding to the interior of the effective two-sided black hole composed of the zone at $t$ and its second exterior mirror.
 To obtain the whole black hole interior, we need to patch the pictures obtained by effective theories erected at different exterior times.
}
\label{fig:eff-int-region}
\end{figure}

From the exterior description, the global spacetime picture is obtained by erecting effective theories of the interior described above at different times $t$ and ``patching'' them~\cite{Nomura:2018kia}, as sketched in Fig.~\ref{fig:eff-int-region}.
We first highlight two important aspects in this procedure:
\begin{itemize}
\item[(i)]
There are many different ways to represent the same interior operator in fundamental (hard, soft, and far) degrees of freedom.
To see this, consider the state at time $t' > t + O(1/T_{\rm H})$ of the system that was described by~\eqref{eq:BH-state} at time $t$.
This state can be written as
\begin{equation}
  \ket{\Psi(M')} \approx \sum_{n'} \sum_{i'_{n'} = 1}^{e^{S_{\rm bh}(M'-E_{n'})}} \sum_{a'} c_{n' i'_{n'} a'} \ket{\{ n'_{\alpha'} \}} \ket{\psi^{(n')}_{i'_{n'}}} \ket{\phi_{a'}},
\label{eq:BH-state-3}
\end{equation}
where $M' < M - O(T_{\rm H})$ is the black hole mass at $t'$, and $\ket{\{ n'_{\alpha'} \}}$, $\ket{\psi^{(n')}_{i'_{n'}}}$, and $\ket{\phi_{a'}}$ are the states of the hard, soft, and far modes \textit{defined at that time}.
If $t' - t$ is smaller than the signal propagating time between the edge of the zone and the stretched horizon, then the interior portions of $K_t$ and $K_{t'}$ constructed from the two states in Eqs.~\eqref{eq:BH-state} and \eqref{eq:BH-state-3} overlap.
This allows us to represent the same interior operator in $K_t \cap K_{t'}$ differently---as $O$ and $O'$---using the microscopic degrees of freedom appearing in two vacuum microstates Eqs.~\eqref{eq:BH-state} and \eqref{eq:BH-state-3}.

For $M - M' > O(T_{\rm H})$, the hard mode states appearing in Eqs.~\eqref{eq:BH-state} and \eqref{eq:BH-state-3} are orthogonal (possibly up to exponentially small overlaps), since the black holes at times $t$ and $t'$ are macroscopically distinguishable; the same is true for the soft and far mode states.
It follows that $O$ and $O'$ are not related simply by the original exterior time evolution $U(t',t)$:
\begin{equation}
  O' \neq U(t',t)\, O\, U^\dagger(t',t).
\end{equation}
This can be seen, for example, from the fact that some of the hard modes at time $t$ evolve into far modes at $t'$, and that for a young black hole, interior operators can be constructed without involving far mode degrees of freedom.
The argument here demonstrates that a version of quantum error correction discussed in~\cite{Almheiri:2014lwa} also applies here in constructing interior operators from the bulk degrees of freedom, i.e.\ the hard, soft, and far modes.
\item[(ii)]
Different operators in the interior region, in particular operators that are independent at the semiclassical level, involve the same microscopic degrees of freedom.
This is, in a sense, the converse of the phenomenon described in (i).
This violation of independence comes from the fact that soft modes at some time $t$ are not necessarily independent from hard modes at some other time $t'$, as they can be related by time evolution $U(t,t')$.
However, since the number of independent soft mode and radiation states is exponentially larger than that of hard mode states, the degree of violation arising in this way is suppressed exponentially by the factor $e^{-S_{\rm tot}/2}$.
\end{itemize}
These two features reinforce the conclusion that the picture of the black hole interior is intrinsically semiclassical: it is defined only up to the precision of order $e^{-S_{\rm tot}/2}$.

We now discuss the third aspect, which is particularly relevant for our discussion:
\begin{itemize}
\item[(iii)]
If the state at time $t$ is given by~\eqref{eq:BH-state}, then the operator $\Phi_\Gamma(x)$ that acts \textit{on this state} and represents a local Heisenberg-picture operator at $x \in K_t$ in~\eqref{eq:K_t} is given by
\begin{equation}
  \Phi_\Gamma(x) = \sum_{s,\Omega,{\bf L}} \left( f_s(\Omega,{\bf L})\, \varphi_{\Omega,{\bf L}}(x)\, a_\xi + g_s(\Omega,{\bf L})\, \varphi_{\Omega,{\bf L}}^*(x)\, a_{\xi^c}^\dagger \right).
\end{equation}
Here, we have decomposed index $\xi$ of the infalling mode operators $a_\xi$ and $a_\xi^\dagger$ into $\Gamma$, $s$, $\Omega$, and ${\bf L} = \{ \ell, m \}$ which represent species, spin, frequency, and orbital angular momentum quantum numbers, respectively, with $\xi^c$ representing the CPT conjugate of $\xi$; $f_s(\Omega,{\bf L})$ and $g_s(\Omega,{\bf L})$ are the standard factors providing Lorentz representation of the field (Dirac spinors, polarization vectors, etc), and $\varphi_{\Omega,{\bf L}}(x)$ are the mode functions.
Note that since $a_\xi$ and $a_\xi^\dagger$ are given by Bogoliubov transformations of $b_\gamma$, $b_\gamma^\dagger$, $\tilde{b}_\gamma$, and $\tilde{b}_\gamma^\dagger$ in Eqs.~\eqref{eq:ann-cre-b} and \eqref{eq:ann-cre-b-tilde}, these operators act on the hard, soft, and far mode states defined at $t$.

As depicted in Fig.~\ref{fig:eff-int-region}, in order to cover the entire black hole interior $I$, we need to erect the effective theory of the interior in at least every time interval of order $1/2\pi T_{\rm H}$.
In other words, in order to obtain the complete set of interior operators $\{ \Phi_\Gamma(x)\, |\, x \in I \}$, we need to use $a_\xi$ and $a_\xi^\dagger$ defined at different times $t_x$ in constructing operators at different locations $x$, since not all $x \in I$ belong to the same $K_t$ for some $t$.
While operators obtained in this way act generally on the hard, soft, and far mode states at different times $t_x$, we can time evolve them to obtain the set of operators that all act on the radiation state at time $t_{\rm r}$ after the black hole is fully evaporated, i.e.\ $t_{\rm r} > t_{\rm evap}$:
\begin{equation}
  {\cal I}_{t_{\rm r}} = \bigl\{ {\cal O}_{\Gamma,x}({t_{\rm r}}) \equiv U(t_{\rm r},t_x) \Phi_\Gamma(x) U^\dagger(t_{\rm r},t_x)\, \big|\, x \in I \bigr\}.
\label{eq:interior-ops}
\end{equation}
As discussed in~\cite{Langhoff:2020jqa,Murdia:2022giv}, this is how entanglement wedge reconstruction~\cite{Czech:2012bh,Wall:2012uf,Headrick:2014cta,Jafferis:2015del,Dong:2016eik,Faulkner:2017vdd,Cotler:2017erl,Akers:2020pmf} represents all the interior operators in the final state Hawking radiation.
We stress that from this perspective, entanglement wedge reconstruction of interior operators from Hawking radiation involves evolution \textit{backward in time}, despite the fact that the interior region and the region in which the radiation exists are spacelike separated in the global spacetime picture.
\end{itemize}

The operators ${\cal O}_{\Gamma,x}({t_{\rm r}})$ acting on final-state Hawking radiation represent local operators in the interior of the black hole, a region \textit{spacelike separated} from the region where the radiation resides.
This seems to imply that an operation on the Hawking radiation corresponding to the application of ${\cal O}_{\Gamma,x}({t_{\rm r}})$ can affect the experience of an observer falling into the black hole, potentially violating the causality of general relativity.%
\footnote{
 This argument can be extended to radiation emitted from an old black hole, i.e.\ to the case $t_{\rm Page} < t_{\rm r} < t_{\rm evap}$, where $t_{\rm Page}$ is the Page time~\cite{Page:1993wv}.
 In this case, the amount of information in the relevant interior region, $(\bigcup_{t < t_{\rm r} - t_{\rm scr}}\! K_t) \cap I$, accessible in the radiation is limited and depends on the location in this region, where $t_{\rm scr}$ is the scrambling time~\cite{Hayden:2007cs,Sekino:2008he}.
 See Fig.~13 of~\cite{Murdia:2022giv} and discussion associated with it for more details.
\label{ft:extension}}
This issue was studied in~\cite{Kim:2020cds} (see also~\cite{Harlow:2013tf,Brown:2019rox}) for radiation before complete evaporation (the case in footnote~\ref{ft:extension}), wherein it was shown that the operation of acting ${\cal O}_{\Gamma,x}({t_{\rm r}})$ is so highly complex that locality in the semiclassical description (i.e.\ the spacelike separated nature between the interior and radiation regions) is preserved for all practical purposes.

However, as we will see in the next section, this leaves a major puzzle associated with causality in the global spacetime picture.
We will see that the construction given in~\cite{Nomura:2018kia,Nomura:2019qps,Nomura:2019dlz,Langhoff:2020jqa,Nomura:2020ska,Murdia:2022giv} and described here addresses this puzzle.
In particular, we will see that in our picture, unlike what was envisioned in earlier works~\cite{Kim:2020cds,Bousso:2022hlz,Bousso:2023sya}, one \textit{cannot} affect the experience of an observer who has fallen into the black hole at an earlier time by performing an (even complex) operation on radiation emitted from the black hole.
This is because, at the microscopic level, the region inside the black hole should be regarded as being \textit{timelike separated} from the region in which the radiation resides; more specifically, the former is in the past of the latter.
The fact that operators like ${\cal O}_{\Gamma,x}({t_{\rm r}})$ exist simply reflects the standard statement that there are operators at a time $t$ corresponding to changing the \textit{record} of what happened at times earlier than $t$.
Together with the fact that the ${\cal O}_{\Gamma,x}({t_{\rm r}})$ effectively commute with a class of simple operators acting on radiation---the feature arising from the involvement of complex horizon dynamics---this allows us to treat the past event represented by ${\cal O}_{\Gamma,x}({t_{\rm r}})$ as occurring in a spacelike separated region in semiclassical spacetime.

\section{A Puzzle in the Global Spacetime Picture and Its Resolution}
\label{sec:puzzle-sol}

In this section, we discuss a problem with global spacetime when it is used to formulate standard quantum mechanics at the fundamental level.
We will then show that by using the description of the black hole interior presented in the previous section, this problem is avoided.
In this description, the Hawking radiation containing information about the interior always lies in the causal future of the corresponding interior region.
It is only in the semiclassical description that the interior appears causally disconnected from the radiation region.

We will discuss in detail how this apparent change of the description arises from the complex dynamics at the horizon, as viewed from the exterior.
By the semiclassical description, we mean the description including only finite products of local operators in the limit that the Newton constant is sent to zero, $G_{\rm N} \rightarrow 0$.
For operators acting on the final-state Hawking radiation, this corresponds to operators that act on a number of degrees of freedom which does not scale with the entropy of the radiation.%
\footnote{
 This implies that the semiclassical theory does not describe \textit{everything} about the low energy degrees of freedom.
}
We will show that these operators all effectively commute with operators describing the interior, so that they can be viewed as spacelike separated at the semiclassical level.

\subsection{A problem with the microscopic description of global spacetime}
\label{subsec:problem}

We argue that the following three postulates cannot be satisfied at the same time:
\begin{itemize}
\item[(i)]
Diffeomorphism invariance persists throughout inextendible (global) spacetime, i.e.\ the region in which the classical description of spacetime is available in general relativity.
\item[(ii)]
The canonical formalism of quantum mechanics can be built using Cauchy surfaces in global spacetime, at least when they satisfy certain ``niceness'' properties~\cite{Lowe:1995ac,Giddings:2006sj,Bousso:2022tdb}.
\item[(iii)]
An observer in the exterior region can reconstruct local operators in the interior using only Hawking radiation degrees of freedom.
\end{itemize}
We will show this by demonstrating that the above three postulates, taken together, lead to a violation of the no-cloning theorem of quantum mechanics.

To see this, consider a setup in which a detector staying outside the horizon throughout the history of the black hole evolution extracts information about the interior at some time $t_{\rm r}$ after the Page time $t_{\rm Page}$.%
\footnote{
 Strictly speaking, $t_{\rm r}$ is not exactly a single time but has a ``width,'' since it will take some time for the detector to extract the information from Hawking radiation through interactions with it; but we will loosely refer to the period of time in which the detector extracts the information as $t_{\rm r}$.
}
For simplicity, we focus on the case in which $t_{\rm r}$ is after the full evaporation of the black hole, $t_{\rm r} > t_{\rm evap}$, though the extension to the case with $t_{\rm Page} < t_{\rm r} < t_{\rm evap}$ is straightforward (see footnote~\ref{ft:extension}).

While the operation performed by the detector to achieve the information extraction needs to be highly nonlocal, acting on most of the radiation degrees of freedom, we can fine-tune the initial configuration of the detector so that it reads off information corresponding to a semiclassical excitation in the interior.%
\footnote{
 The corresponding operator acting on the radiation comprises a product of $S_{\rm rad}$ local operators, where $S_{\rm rad}$ is the coarse-grained entropy of the radiation, so that the required fine-tuning is of order $e^{-O(S_{\rm rad})}$.
 We do not consider such an operator to be a part of the semiclassical theory.
 Thus, the problem discussed here is not that of the semiclassical description but of the fundamental description applied to the global spacetime.
}
As a concrete example, we can set up the initial configuration of the detector such that it interacts with the degree of freedom in the Hawking radiation which is represented by a product of operators ${\cal O}_{\Gamma,x}({t_{\rm r}})$ for some spacetime point $x \in I$.
Note that this requires an exponential fine-tuning of the configuration of the detector, which must be determined by the microscopic theory using the knowledge of the initial state of collapsing matter forming the black hole.
This is difficult in practice but possible in principle, because we can have as much time as we need to prepare the detector before actually forming the black hole.

\begin{figure}[t]
\centering
  \includegraphics[height=0.49\textwidth]{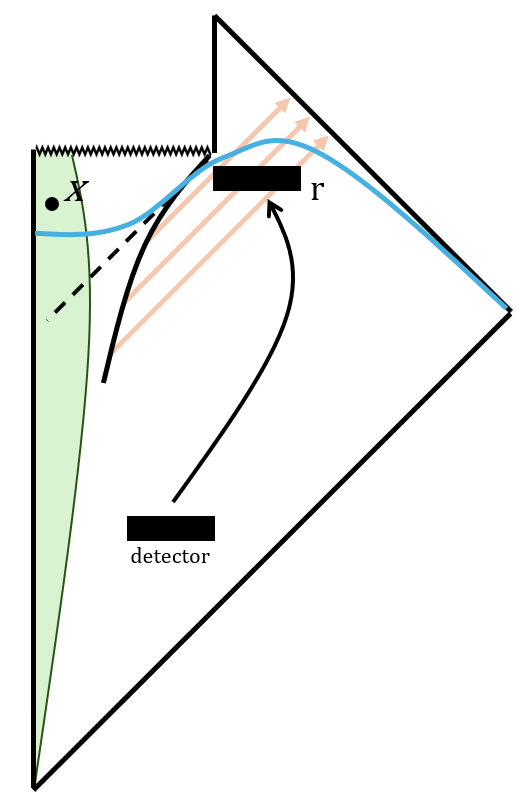}
\vspace{2mm}
\caption{
 A detector can be set up such that it extracts from Hawking radiation the information that is located at a spacetime point $x$ in the interior of the black hole.
 Interpreting this as being caused by nonlocal interactions between $x$ and the spacetime region $r$ in which the radiation resides, we encounter the problem of the existence of a time slice (blue line) that goes through both the future of $r$ and the past of $x$.
 On such a slice, the information in question is cloned, leading to a contradiction with the no-cloning theorem of quantum mechanics.
}
\label{fig:clone-slice}
\end{figure}

With this setup, we find it problematic to consider the global spacetime as a platform for a regular quantum mechanical system.
In particular, we cannot employ the canonical formalism of quantum mechanics using a series of Cauchy surfaces, such as nice slices~\cite{Lowe:1995ac,Giddings:2006sj,Bousso:2022tdb}, in which the information extracted by the detector is removed from the future of $x$ in the interior and appears in the future of $r$ in the exterior region.
This can be seen by considering a foliation of spacetime accommodating a time slice that goes through both the past of $x$ and the future of the region in which the detector extracts the information; see Fig.~\ref{fig:clone-slice}.
On this slice, the information extracted by the detector is cloned, since the destruction of the same information in the interior occurs at $x$ and hence has not yet happened on this time slice.
This contradicts the no-cloning theorem, hence violating the superposition principle of quantum mechanics.
We stress that this problem occurs even when we only use slices that are nice in the semiclassical description.

One might think that this problem is avoided by claiming that the microscopic description of the global spacetime forces us to use a special foliation in which the time slice going through $t_{\rm r}$ \textit{must} also go through $x$.
We find it difficult, however, to make sense of such a claim given that the arbitrariness of choosing time foliation, i.e.\ diffeomorphism invariance, plays an important role in ensuring the consistency of theories with dynamical gravity, even at the perturbative level.%
\footnote{
 This is a difficulty which we encounter in addition to the fact that the time $t_{\rm r}$ has a width, so that the relevant time slice would generally be ill-defined.
}

\subsection{A solution:\ the interior as an effective description of dissipation on the horizon}
\label{subsec:solution}

The above argument implies that, if we are to keep~(iii), we must relinquish~(i) or (ii).
The understanding of the black hole interior described in Section~\ref{sec:emergence} addresses the puzzle by abandoning postulate~(ii).
In this framework, the interior spacetime---in fact only a portion of it---emerges as an effective description associated with a given exterior time.
The effective time evolution foliating the interior region is different from the more fundamental, exterior time evolution, which foliates only the spacetime region outside the horizon.
Consequently, any information in the effective interior is contained in modes \textit{outside the horizon}, and hence exists in the causal past of the radiation.
By viewing the interior this way, we need not demand that the interior is foliable by global time slices, and hence do not take postulate~(ii).
The property in postulate~(i) emerges only effectively after patching the pictures obtained by effective theories erected at different times.
This results in postulate~(iii) not implying that an operation on Hawking radiation can affect the physics in the interior, avoiding the problem described in Section~\ref{subsec:problem}.

\begin{figure}[t]
\centering
  \includegraphics[height=0.3\textheight]{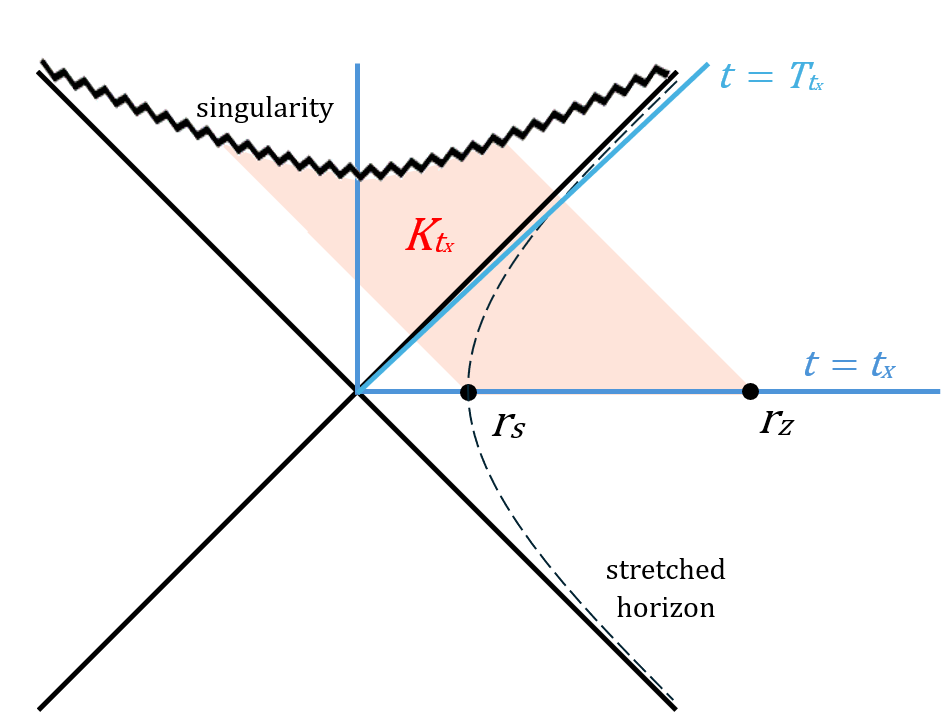}
\vspace{1mm}
\caption{
 The physics occurring in the region described by the effective theory erected at time $t_x$ cannot be obtained by any effective theory erected after $T_{t_x} = t_x + 2r_+ \ln(r_+/l_{\rm P})$, implying that it must be viewed as having occurred before $T_{t_x}$ in the exterior description.
}
\label{fig:t_eff-x}
\end{figure}

Recall that in the framework of Section~\ref{sec:emergence}, physics occurring at the interior point $x$ is described by an effective theory erected at a time $t_x$ ($< t_{\rm evap}$) satisfying
\begin{equation}
  x \in K_{t_x},
\label{eq:t_x}
\end{equation}
where $K_t$ is given by~\eqref{eq:K_t}.%
\footnote{
 It is important that $K_{t_x}$ only contains the future of the region \textit{outside} the stretched horizon; it does not contain the region which is in the future of $r < r_{\rm s}$ at $t_x$ but not in the future of $r > r_{\rm s}$ at $t_x$, as illustrated in Fig.~\ref{fig:t_eff-x}.
 This ultraviolet cutoff is a manifestation of the fact that the effective theory of the interior is intrinsically semiclassical, without which all the radial null geodesics entering the stretched horizon before $t_x$ would pile up along the horizon of the second exterior in the effective theory causing large backreaction on the spacetime~\cite{Nomura:2018kia}.}
As can be seen in Fig.~\ref{fig:t_eff-x}, the spacetime region $K_{t_x}$ described by the effective theory contains (a portion of) the zone, $r_{\rm s} < r < r_{\rm z}$, only up to the time
\begin{equation}
  T_{t_x} = t_x + 2r_+ \ln\frac{r_+}{l_{\rm P}},
\end{equation}
where $r_+$ is the horizon radius of the black hole \textit{at time $t_x$}, and we have ignored the term of order $r_+$ which is not enhanced by the logarithm $\ln(r_+/l_{\rm P})$.
(We will also do the same below.)
This implies that in the exterior description, the physics occurring in the interior portion of $K_{t_x}$ should be viewed as occurring in the time frame
\begin{equation}
  t_x \,<\, t \,<\, T_{t_x},
\label{eq:eff-time}
\end{equation}
since the effective theory simply reorganizes correlations between the degrees of freedom associated with the time slices in~\eqref{eq:eff-time} such that they can be interpreted as arising from the infalling time evolution generated by $\tilde{H}$ in~\eqref{eq:tilde-H}, \textit{without} involving a further exterior time evolution.
The earliest effective theory to which $x$ can be associated is the one erected at time $t_{\rm min} = \text{min}\{t_x : x\in K_{t_x}\}$.
The latest effective theory which $x$ can be associated with is similarly the one erected at $t_{\rm max} = \text{max}\{t_x : x\in K_{t_x}\}$.
Consequently, $x$ must be associated with exterior modes between $t_{\rm min}$ and $T_{t_{\rm max}}$.
In other words, the physics occurring in the interior spacetime point $x$ must be viewed as occurring in the time frame 
\begin{equation}
  t_{\rm min} \,<\, t \,<\, T_{t_{\rm max}}
\label{eq:eff-time-2}
\end{equation}
in the exterior description.

Importantly, for any point $x$ in the interior of the black hole
\begin{equation}
  T_{t_{\rm max}} < t_{\rm evap}.
\end{equation}
This implies that in the exterior description, the physics which an interior observer experiences at the spacetime point $x$ must have \textit{already happened} before the black hole evaporates.
Thus, whatever the detector does to the final state Hawking radiation, it \textit{cannot affect} their experience at $x$ nor events occurring in its future.
Exerting operators ${\cal O}_{\Gamma,x}({t_{\rm r}})$ on the radiation at $t_{\rm r}$ can \textit{reconstruct} the event at $x$, or even \textit{rewrite the record} of what happened at $x$ and in its future; however, it cannot change the actual experience which the interior observer has had inside the black hole.

In this respect, the situation is not so different from the standard situation in which a complicated operation can be applied to the system ``after the fact'' so that it can change the record of what has happened before the operation.%
\footnote{
 Throughout the paper, we assume that things happen when they actually occur through time evolution.
 In other words, we consider that a measurement occurs when the relevant information corresponding to the measurement outcome (which is not full quantum information) is amplified into various subsystems~\cite{Nomura:2011rb,q-Darwinism,q-Darwinism-2}.
 In particular, we do not take an extreme (and what we think an exotic) standpoint~\cite{Bousso:2011up} that a measurement occurs only when the relevant information reaches to a boundary or horizon region.
}
In our setup, if we reconstruct from Hawking radiation what happened in the interior of the black hole before applying the complex operation, i.e.\ at $t < t_{\rm r}$, then we would conclude that nothing extraordinary happened at $x$ in the interior.
On the other hand, if we do such a reconstruction after the detector exerts the complex operation, i.e. at $t > t_{\rm r}$, then we would conclude that an event occurred at $x$ ``out of the blue'' which did affect the future of $x$ in the interior.
(The past of $x$ is unaffected because the boundary condition for the interior is determined by the exterior state at $t_x$, fixing the state in the past light cone of $x$.)
This ``nonlocal influence'' by the operation, however, is an illusion---what a physical observer \textit{actually experiences} in the interior is not affected by the operation of the detector.

In the exterior description, the event $x$ must thus be viewed as having occurred \textit{in the past of} $r$, despite the fact that they are spacelike separated in the semiclassical description.
This picture becomes evident because we have employed the construction in~\cite{Nomura:2018kia,Nomura:2019qps,Nomura:2019dlz,Langhoff:2020jqa,Nomura:2020ska,Murdia:2022giv}, which makes manifest the dynamical nature of interior reconstruction, i.e., how the exterior time evolution is involved in the reconstruction.
In fact, this feature is not visible in a model without appropriately taking into account time evolution, for example in the case of a tensor network corresponding to a single global time slice.

An alternative way of viewing all of this is that to correctly reconstruct the interior, we need to include the agent, e.g.\ a detector, which has exerted operators ${\cal O}_{\Gamma,x}({t_{\rm r}})$ on the radiation.
We should then evolve the \textit{combined} system of the agent and radiation backward in time and use the construction of~\cite{Nomura:2018kia,Nomura:2019qps,Nomura:2019dlz,Langhoff:2020jqa,Nomura:2020ska,Murdia:2022giv} at time $t_x$. By doing this, rather than simply adopting the entanglement wedge reconstruction protocol to the radiation after ${\cal O}_{\Gamma,x}({t_{\rm r}})$'s are operated, we will indeed reconstruct the interior in which nothing extraordinary happens at $x$.
The inclusion of the agent affects the interior significantly, despite the fact that it comprises a much smaller number of degrees of freedom than those of the radiation.
This is because the standard process of black hole formation and evaporation is one in which the coarse-grained entropy increases~\cite{Zurek:1982zz,Page:1983ug}, and hence the reverse time evolution employed during interior reconstruction is a highly fine-tuned, fragile process.

\begin{figure}[t]
\centering
  \includegraphics[height=0.34\textheight]{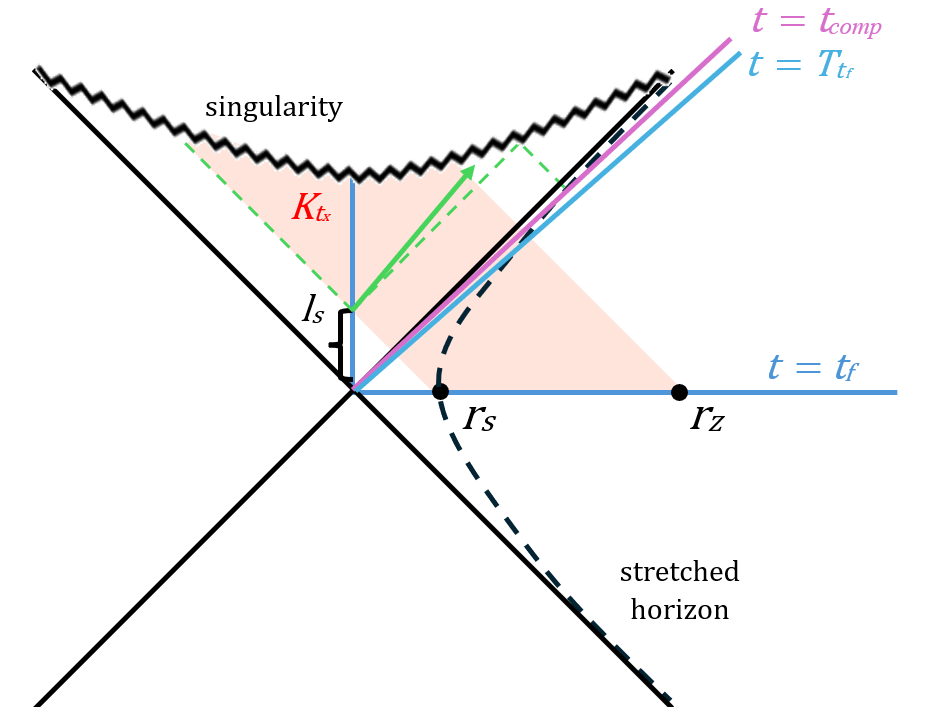}
\vspace{1mm}
\caption{
 The physics associated with an object that fell into the horizon at time $t_{\rm f}$ cannot be described by any effective theory erected after $t_{\rm comp} = t_{\rm f} + t_{\rm scr}$.
 This implies that it must be viewed as having occurred before $t_{\rm comp}$ in the exterior description.
}
\label{fig:t_scr}
\end{figure}

The analysis here can be extended to the case in which we act operators ${\cal O}_{\Gamma,x}({t_{\rm r}})$ before the black hole is fully evaporated, which is possible if $t_{\rm r} > t_{\rm Page}$.
To see this, consider an infalling observer who reaches the stretched horizon at some exterior time $t_{\rm f}$.
While the experience of the observer shortly after crossing the horizon is described by the effective theory of the interior erected at $t_{\rm f}$, this does not necessarily mean that all the experiences of this observer are described in this effective theory.
For example, the observer may accelerate toward the horizon right after they cross it, eventually leaving $K_{t_{\rm f}}$; see Fig.~\ref{fig:t_scr}.
Even in this case, however, the experience of the infallen observer should be viewed as fully completed by
\begin{equation}
  t_{\rm comp} = t_{\rm f} + t_{\rm scr}
\end{equation}
in the exterior description, where
\begin{equation}
  t_{\rm scr} = 4r_+ \ln\frac{r_+}{l_{\rm P}}
\end{equation}
is the scrambling time.
One way to see this is to notice that the late trajectory of the observer is fully contained in the effective theory erected at $t'_{\rm f} = t_{\rm f} + 2r_+ \ln(r_+/l_{\rm P})$, even in the worst case that the observer starts moving outward at the speed of light one cutoff time after crossing the mathematical horizon.
This is indicated by the arrow in Fig.~\ref{fig:t_scr}.
The argument that led to~\eqref{eq:eff-time} then says that any experience of the observer in the interior must have occurred by
\begin{equation}
  T_{t'_{\rm f}} = t_{\rm f} + t_{\rm scr}.
\end{equation}
On the other hand, the general argument in~\cite{Hayden:2007cs} (or entanglement wedge reconstruction~\cite{Penington:2019npb}) says that the information about an object that fell into the black hole at $t_{\rm f}$ emerges in Hawking radiation only after $t_{\rm f} + t_{\rm scr}$.
Thus, again any operation on Hawking radiation---even if it is performed before $t_{\rm evap}$, and no matter how complex it is---cannot affect the experience of an observer in the interior.
This conclusion holds even if we mine~\cite{Unruh:1982ic,Brown:2012un} radiation from the Hawking cloud in the zone around the black hole.%
\footnote{
 To reconstruct the interior of the black hole at an earlier time $t$ from radiation at $t_{\rm r}$ ($> t$, $< t_{\rm evap}$), we need to include all semiclassically distinguishable branches (with different black hole masses and positions) that have resulted from the backreaction~\cite{Page:1979tc,Nomura:2012cx} of Hawking emission occurring after $t$.
}

Finally, our analysis yields the following interpretation~\cite{Nomura:2018kia} of the black hole's singularity.
From an exterior perspective, an object does not immediately thermalize when it encounters the stretched horizon.
Instead, it manifests as a distinct excitation on the horizon for some duration.
Over the scrambling time, $t_{\rm scr}$, this excitation will gradually dissipate into the horizon. 
During this period, the dissipation process has an effective description as an object traversing the near-empty interior of the black hole.
Here, coherent ``quasi-particle'' horizon modes---and beyond the Page time, early radiation modes as well---denoted by $\tilde{b}_\gamma$, serve as modes in the effective second exterior of the black hole.
The singularity of the semiclassical picture signifies the limit of the applicability of this effective description.
After the object's excitation is fully integrated into the horizon, the effective theory describing the dissipation process is no longer available.

\subsection{Converting relations from timelike to spacelike using complexity}
\label{subsec:time-space}

It is highly nontrivial to regard two sets of degrees of freedom that are not independent with each other as spacelike separated at the semiclassical level.
If interior reconstruction indeed works, Hawking radiation for an old black hole is not independent of the interior degrees of freedom. 
To regard these as spacelike separated, while preserving causality at the semiclassical level, an operator acting on one set must commute with that acting on the other, possibly up to corrections suppressed exponentially in a macroscopic entropy.
Here we argue that our framework indeed implies that a process involving only low energy degrees of freedom cannot affect the semiclassical physics in a region that can be viewed as spacelike separated.

In~\cite{Kim:2020cds}, Kim, Tang, and Preskill (KTP) analyzed this issue of causality, assuming that the black hole's dynamics can be represented by pseudorandom unitaries~\cite{Kim:2020cds,Bouland:2019pvu,Engelhardt:2024hpe}.
Since it is expected that chaotic dynamics generally produce pseudorandom states sufficiently after the scrambling time, we assume that the same applies to the dynamics of our horizon modes as viewed from the exterior.
KTP studied a partially evaporated black hole and concluded that if an operator ${\cal O}_{\rm rad}$ acting on the radiation can be modeled by a quantum circuit with size polynomial in the entropy of the black hole, i.e.\ ${\cal C}({\cal O}_{\rm rad}) \sim {\rm poly}(S_{\rm BH})$, then ${\cal O}_{\rm rad}$ commutes with operators ${\cal O}(x)$ describing the interior up to corrections suppressed exponentially in $S_{\rm BH}$
\begin{equation}
  \left\lVert [{\cal O}_{\rm rad}, {\cal O}(x)] \right\rVert \sim O(e^{-S_{\rm BH}}).
\end{equation}
Applying the same argument to the final-state Hawking radiation, we could conclude that an operator ${\cal O}'_{\rm rad}$ acting only on a portion $S'_{\rm rad}$ ($< S_{\rm rad}$) of the radiation whose complexity goes as
\begin{equation}
  {\cal C}({\cal O}'_{\rm rad}) \sim {\rm poly}\left( S_{\rm rad} - S'_{\rm rad} \right)
\end{equation}
would effectively commute with operators describing the interior of the black hole:\ $\lVert [{\cal O}'_{\rm rad}, {\cal O}(x)] \rVert \sim O(e^{-(S_{\rm rad} - S'_{\rm rad})})$.

However, this by itself does not solve the issue, since one would still be able to act on an operator on the entire final-state radiation ($S'_{\rm rad} \approx S_{\rm rad}$) and affect the interior; indeed in that case the operator can be simple.%
\footnote{
 Here we have assumed, following~\cite{Kim:2020cds}, that the state of final-state Hawking radiation is efficiently generated.
 It seems possible to us that the generation of final-state Hawking radiation could be hard, which can happen if, for example, the computational basis of Hawking radiation (the basis determined by spatial locality) and the natural basis of the horizon dynamics are related by a generic unitary.
 If this is the case, then the operator affecting the interior must be complex, with the complexity ${\cal C}$ scaling exponentially in $S_{\rm rad}$.
}
This would lead to the problem discussed in Section~\ref{subsec:problem}.
As we have seen, however, the Hawking radiation containing information about the interior is, in fact, \textit{timelike} related with the interior degrees of freedom.
The former is in the future of the latter.
Therefore, one cannot change the physics that has already occurred in the interior by acting with an operator on the Hawking radiation, \textit{no matter how complex it is}.

This leaves the question about the effect of an interior operator (past) on Hawking radiation (future).
In order to preserve causality, we want the semiclassical physics associated with the radiation not to be affected by an operation in the interior.
By semiclassical physics, we mean the collection of processes represented by operators which act on degrees of freedom whose numbers do not scale with $S_{\rm rad}$, which is proportional to the initial entropy of the black hole and hence to $1/G_{\rm N}$.
To study this question, we can employ a variant of the KTP analysis; the interior operator in KTP is replaced with a semiclassical operator acting on the radiation, and the combined system of the black hole and radiation in KTP is replaced with the black hole degrees of freedom at the time when the effective theory of the interior is erected.%
\footnote{
 The relation between the degrees of freedom representing the interior and Hawking radiation always involves the horizon dynamics, which we assume to be pseudorandom.
 The analysis of KTP can thus be adopted in the present context.
}
The number of degrees of freedom representing semiclassical physics in the interior is much smaller than that representing the black hole, $S_{\rm BH}$.
The KTP argument then implies that an interior operator ${\cal O}_{\rm int}$ corresponding to a circuit of size
\begin{equation}
  {\cal C}({\cal O}_{\rm int}) \sim {\rm poly}\left( S_{\rm BH} \right)
\end{equation}
effectively commutes with semiclassical operators $\tilde{\cal O}_{\rm rad}$ acting on the radiation, up to corrections suppressed exponentially in $S_{\rm BH}$.
\begin{equation}
  \left\lVert [{\cal O}_{\rm int}, \tilde{\cal O}_{\rm rad}] \right\rVert \sim O(e^{-S_{\rm BH}}).
\end{equation}
In other words, in order to affect the semiclassical physics of the radiation, an operation performed in the interior must have complexity exponential in $S_{\rm BH}$.
This is, however, impossible because any physical ``experimenter'' in the interior hits the singularity in polynomial time before any such operation can be performed.
Note that we have not imposed any constraints on the computational complexity of $\tilde{\cal{O}}_{\rm rad}$; we have just imposed the restriction on the number of degrees of freedom on which it acts.

Any physical operation in the black hole interior, therefore, cannot affect semiclassical physics occurring in the radiation at later times, except for corrections suppressed exponentially in the entropy of the black hole.
The only operators acting on the radiation which do not commute with interior operators are those acting on a large (macroscopic) number of degrees of freedom scaling as $S_{\rm rad}$.
Of course, these noncommutativities are what allow us to decode the information about the interior from the Hawking radiation.

Finally, we note that the analogous statement cannot be made for reverse time evolution.
We \textit{can} perform a physical operation on the (now initial-state) radiation which affects semiclassical physics in the interior (of the white hole).
In this sense, we may say that the Penrose diagram of Fig.~\ref{fig:exterior-slice} faithfully represents physics for the standard black hole evolution, a process in which the coarse-grained entropy increases, but not for the reverse evolution, a process with decreasing coarse-grained entropy caused by a finely-tuned initial condition.

\section{Causal Structure of Black Hole Formation and Evaporation}
\label{sec:causal}

\begin{figure}[t]
\centering
  \includegraphics[height=0.37\textheight]{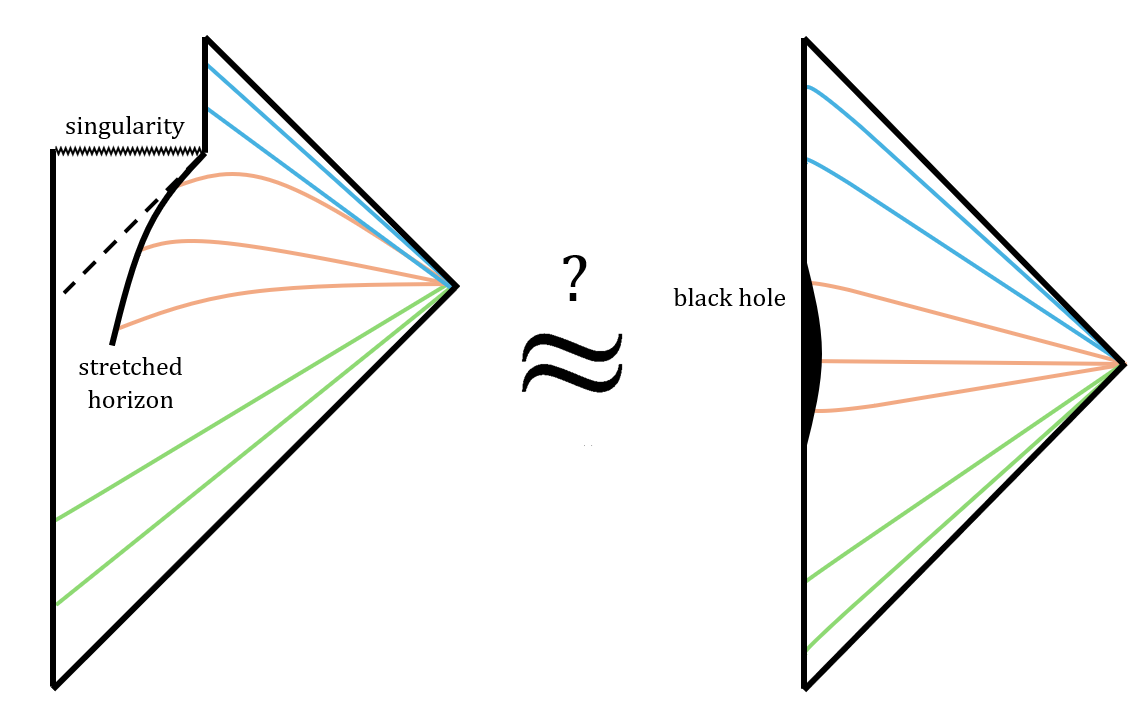}
\vspace{-1mm}
\caption{
 In the exterior description, the causal flow of information is obvious after the black hole is formed.
 The question is if there exists an analogous picture throughout the history of the black hole evolution, including the times during its formation.
}
\label{fig:causal}
\end{figure}

One virtue of the exterior picture is that the causal flow of information is clear.
When an object falls into a black hole, its information stays on the stretched horizon, whose spacetime trajectory is timelike, and later comes back to ambient space in the form of Hawking radiation.
One may thus wonder if the causal structure of the entire spacetime can in fact be trivial in the exterior description (or at least in some version of the exterior description).
We depict how such a description would look in the right panel of Fig.~\ref{fig:causal}.

\begin{figure}[t]
\centering
  \includegraphics[height=0.39\textheight]{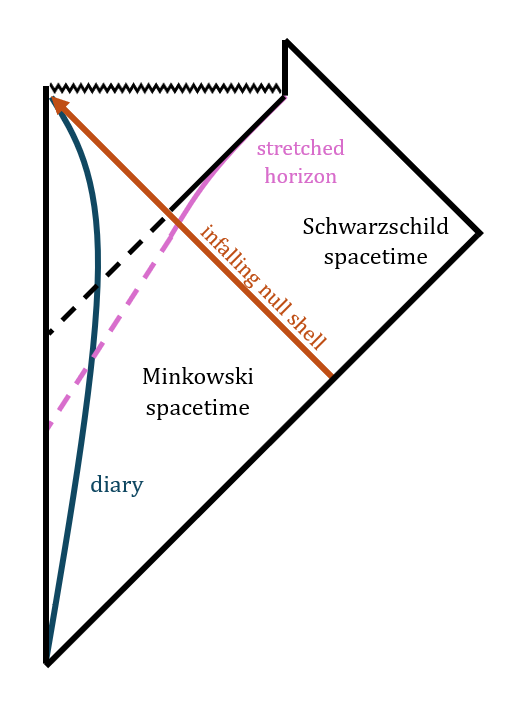}
\vspace{-3mm}
\caption{
 Consider a diary in the region near $r = 0$ far in the past.
 We then fire a null shell of infalling particles toward $r = 0$, creating a black hole of mass $M$.
 The unitarity of the $S$-matrix implies that the information about the diary must eventually be emitted from the stretched horizon in the form of Hawking radiation.
 An exterior description with trivial causality, the right panel of Fig.~\ref{fig:causal}, is obtained by extending the stretched horizon toward the past using the event horizon.
}
\label{fig:formation}
\end{figure}

This would indeed be the case if the ``stretched horizon,'' at which the spacetime terminates, follows a timelike trajectory \textit{throughout} the history of the black hole evolution, including during formation.
Taking this possibility seriously, however, raises an interesting question.
To see this, imagine that a diary has existed at $r = 0$ for an extended period of time in near-empty spacetime.
Then, we fire a null shell of particles toward $r = 0$, creating a black hole of mass $M$; see Fig.~\ref{fig:formation}.
Before the null ray, the spacetime is Minkowski.
After the null ray, the spacetime is Schwarzschild.
Given the unitarity of the $S$-matrix, the information about the diary must eventually be emitted from the stretched horizon in the form of Hawking radiation.
In order to avoid a mysterious spacelike motion of the information from the interior to the exterior, the timelike trajectory of the stretched horizon must extend to the spacetime region before the null ray, as depicted by a curved dashed line in Fig.~\ref{fig:formation}.
However, the spacetime in this region is Minkowski, and there is no obvious locally defined surface which can play the role of the stretched horizon.

\subsection{Exterior description with trivial causal structure}
\label{subsec:trivial-causal}

We argue that the description corresponding to the right panel of Fig.~\ref{fig:causal}, however, is still possible.
In particular, we will provide below a surface which plays the role of the stretched horizon before the null shell, which in the case of Fig.~\ref{fig:formation} is the appropriately stretched horizon of a spherical Rindler observer.
Before the black hole's formation, the information about the diary stays in modes near this horizon, and it will eventually come out in a causal way in the form of Hawking radiation.
This is possible because in quantum gravity the spatial location of information is not uniquely determined.

In quantum gravity, the physics associated with some spacetime region can be encoded in the degrees of freedom on the boundary surrounding it, if certain conditions are met~\cite{tHooft:1993dmi,Susskind:1994vu,Bousso:2002ju}.
In particular, for a closed codimension-2 surface $\Sigma$, the physics occurring in the domain of dependence of a partial Cauchy surface bounded by $\Sigma$ can fully be described using the intrinsically quantum gravitational degrees of freedom on $\Sigma$ and the degrees of freedom outside of it.

We apply this to each Cauchy surface before the formation of the black hole, with $\Sigma$ identified as the (stretched) \textit{event horizon}.
Specifically, in the setup of Fig.~\ref{fig:formation}, we may adopt the ingoing Vaidya metric
\begin{equation}
  ds^2 = -\left( 1 - \frac{2M(v)}{r} \right) dv^2 + 2 dv dr + r^2 d\Omega^2,
\end{equation}
where
\begin{equation}
  M(v) = \left\{ \begin{array}{ll} 0 & \mbox{ for } v < v_0 \\ M & \mbox{ for } v > v_0 \end{array}, \right.
\end{equation}
with $v_0$ the ingoing time at which the null shell is shot.
For $v > v_0$, this corresponds to time slices of the exterior description, as can be seen from the fact that the coordinate transformation
\begin{equation}
  v = t + r + 2M l_{\rm P}^2 \ln\frac{r-2M l_{\rm P}^2}{2M l_{\rm P}^2}
\end{equation}
gives the Schwarzschild metric.
For $v < v_0$, on the other hand, the metric also covers a region inside the event horizon:\ by the coordinate transformation
\begin{equation}
  v = t + r,
\end{equation}
we obtain the flat space metric
\begin{equation}
  ds^2 = -dt^2 + dr^2 + r^2 \Omega^2.
\label{eq:flat}
\end{equation}
The trajectory of the event horizon in this metric is given by
\begin{equation}
  r = t + 4M l_{\rm P}^2 - v_0
\label{eq:event-hor}
\end{equation}
for $v_0 - 4M l_{\rm P}^2 \leq t \leq v_0 - 2M l_{\rm P}^2$.

We may convert the metric~\eqref{eq:flat} into that corresponding to the ``exterior picture,'' by adopting the spherical Rindler coordinates~\cite{Balasubramanian:2013rqa} tailored to the present setup
\begin{equation}
  \tau = {\rm arctanh}\frac{t-t_*}{r},\
\qquad
  \rho = \sqrt{r^2-(t-t_*)^2},
\end{equation}
where
\begin{equation}
  t_* = v_0 - 4M l_{\rm P}^2
\end{equation}
is the time at which the event horizon forms.
This gives the metric
\begin{equation}
  ds^2 = -\rho^2 d\tau^2 + d\rho^2 + \rho^2 \cosh^2\!\tau\, d\Omega^2,
\label{eq:sph-Rindler}
\end{equation}
applicable in the region $v < v_0$, i.e.\ $\rho < 4M l_{\rm P}^2 e^{-\tau}$, and $\tau \geq 0$.
The Rindler horizon agrees with the event horizon~\eqref{eq:event-hor}, which connects smoothly with the black hole horizon at $v = v_0$.
While the system is not in the global thermal equilibrium as the metric~\eqref{eq:sph-Rindler} is not static, observers at constant $\rho$ follow trajectories of constant acceleration and see Unruh radiation~\cite{Unruh:1976db,Fulling:1972md,Davies:1974th}.
The stretched horizon can thus be defined to be located where the local (Tolman) Unruh temperature $1/2\pi\rho$ becomes the string scale $1/2\pi l_{\rm s}$~\cite{Langhoff:2020jqa,Parikh:2017aas}:
\begin{equation}
  \rho = \rho_{\rm s} \approx l_{\rm s}.
\end{equation}
This surface connects smoothly with the stretched horizon of the black hole at $v = v_0$.
At the time of the formation, the area of the stretched horizon is of the cutoff-scale size, ${\cal A}(0) \approx 4\pi l_{\rm s}^2$.
The nonstretched and stretched horizons are depicted by dashed lines in Fig.~\ref{fig:formation}.

We claim that when the theory is quantized canonically, there is a description which keeps only the portion of the spacetime outside the stretched horizon, $\rho > \rho_{\rm s}$, and in which the stretched horizon carries the coarse-grained entropy (the number of effective degrees of freedom)
\begin{equation}
  S_{\rm h}(\tau) = \frac{{\cal A}(\tau)}{4 l_{\rm P}^2},
\label{eq:S_h}
\end{equation}
where
\begin{equation}
  {\cal A}(\tau) = 4\pi \rho_{\rm s}^2 \cosh^2\!\tau
\end{equation}
is the area of the stretched horizon at time $\tau$ ($\geq 0$).%
\footnote{
 The calculation of~\cite{Balasubramanian:2013rqa} gives the von~Neumann entropy, which is zero in the present setup.
}
When the stretched horizon intersects the null shell at
\begin{equation}
  \tau = \ln\frac{v_0-t_*}{\rho_{\rm s}} = \ln\frac{4Ml_{\rm P}^2}{\rho_{\rm s}},
\end{equation}
$S_{\rm h}(\tau)$ and ${\cal A}(\tau)$ agree with the entropy and the area of the stretched horizon of the black hole of mass $M$, respectively.
We postulate that, as suggested by the holographic principle, the degrees of freedom associated with the stretched horizon, represented by~\eqref{eq:S_h}, encode semiclassical physics in the domain of dependence of a partial Cauchy surface bounded by the stretched horizon at time $\tau$
\begin{equation}
  D(\Xi),\qquad \Xi:\, r < \rho_{\rm s} \cosh\tau,\, t = t_* + \rho_s \sinh\tau.
\label{eq:D-Xi}
\end{equation}
Thus, when the diary hits this past-extended stretched horizon, its information is stored in these degrees of freedom.
This information will then be transferred to the black hole horizon degrees of freedom when the stretched horizon is hit by the incoming null shell.

\begin{figure}[t]
\centering
  \subfloat{{\includegraphics[width=0.4\textwidth]{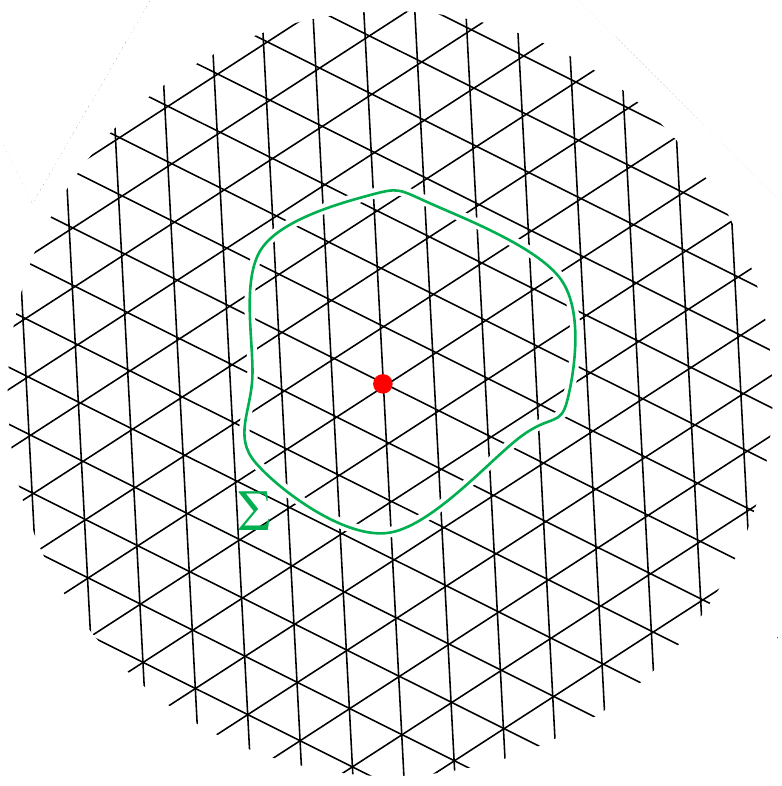} }}
\hspace{1cm}
  \subfloat{{\includegraphics[width=0.43\textwidth]{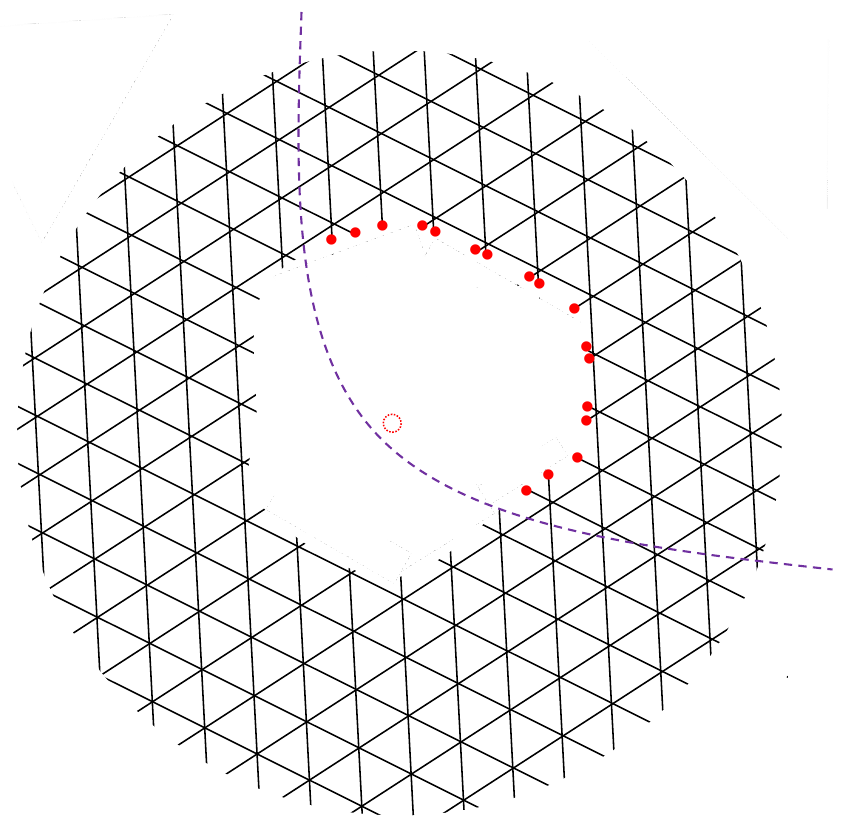} }}
\vspace{3mm}
\caption{
 An excitation in a semiclassical theory can be viewed as a perturbation on the network of entanglement representing a spatial slice (left).
 The network can be cut at a closed surface $\Sigma$ surrounding the perturbation and be replaced with an equivalent network in which the interior of $\Sigma$ is removed (right).
 An operator corresponding to the original excitation can be represented by a nonlocal operator acting on the internal legs of the new network, obtained by pushing the original operator through the network.
 }
\label{fig:tensor}
\end{figure}

To develop intuition about our picture, we can employ a tensor network model~\cite{Swingle:2009bg,Pastawski:2015qua,Hayden:2016cfa}.
In a tensor network, a semiclassical excitation, in this case the diary, can be regarded as a perturbation on the network.
We can insert this perturbation on a specific location in the network, but the same information can then be ``pushed'' through the network and be viewed as existing in external legs of the network, which typically are identified as the boundary degrees of freedom in holography.
The location of this information, however, need not be either the original location of the perturbation or the external legs of the network.
In particular, we can cut the network into two at a closed surface $\Sigma$ which encloses the original location of the perturbation; see Fig.~\ref{fig:tensor}.
The information about the perturbation can then be viewed as existing on legs that go through $\Sigma$.
In fact, if all the information about the inner portion of the network can be represented in this way on the cut legs of the outer portion, then we need not even keep the inner portion to describe the system.

Our description conforms with the recent proposal of Bousso and Penington~\cite{Bousso:2022hlz,Bousso:2023sya} which extends the concept of entanglement wedge reconstruction to bulk subregions.
According to their criterion, the max-entanglement wedge $e_{\rm max}$~\cite{Bousso:2023sya} of the partial Cauchy surface outside the extended stretched horizon at any time is the entire spacetime.
This is consistent with our claim, given that entanglement wedge reconstruction assumes the full knowledge about the (exterior) time evolution, so that it can reconstruct the region $D(\Xi)$~\eqref{eq:D-Xi} for all $t$.

\begin{figure}[t]
\centering
  \includegraphics[height=0.35\textheight]{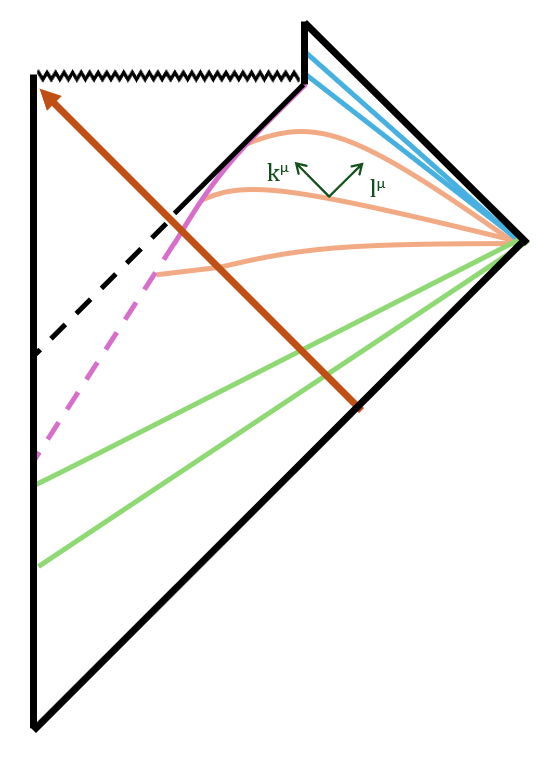}
\vspace{0mm}
\caption{
 An equal-time hypersurface of the holographic boundary, called a leaf, can be pulled inward by a renormalization flow.
 The trajectory of each leaf, depicted in the figure, then provides an equal-time hypersurface in the bulk called a holographic slice in~\cite{Nomura:2018kji,Murdia:2020iac}.
 We can terminate the flow when the renormalized leaf reaches the stretched event horizon, giving the manifestly causal exterior description.
}
\label{fig:flow}
\end{figure}

Our description may also be obtained by ``pulling in'' the boundary of the (putative) holographic theory~\cite{Nomura:2018kji,Murdia:2020iac,Miyaji:2015yva}.
Imagine that there is a ``screen'' in the asymptotic region $r \rightarrow \infty$ on which the holographic theory lives.
(In asymptotically AdS spacetime, this can be the conformal boundary.)
Each time slice of the screen is called a leaf.
We can then renormalize the leaf at each time, which moves the leaf inward in the bulk.
As discussed in~\cite{Murdia:2020iac}, when doing this at the quantum level, we need to keep the information outside the renormalized leaf $\sigma(\lambda)$, where $\lambda$ is the flow parameter.
Specifically, the flow is given by the equation
\begin{equation}
  \frac{dx^\mu}{d\lambda} =\frac{1}{2}(\Theta_k l^\mu+\Theta_l k^\mu),
\end{equation}
where $x^\mu$ are the embedding coordinates of the renormalized leaf $\sigma$, and $\{ k^\mu, l^\mu \}$ are the future-directed null vectors orthogonal to $\sigma$.
Here, $\Theta_{k,l}$ are the modified versions~\cite{Murdia:2020iac} of the quantum expansions~\cite{Bousso:2015mna} in the corresponding directions, whose computation requires knowledge about the state outside $\sigma$.
In the Schwarzschild region the flow terminates near the horizon, but in the Minkowski region it can go all the way toward $r = 0$.
Our description can, thus, be viewed as the holographic theory in which each leaf is renormalized down to the extended stretched horizon $r = r_{\rm s}(t)$ in the Minkowski region.
The region $r < r_{\rm s}(t)$ is then obviously encoded in the renormalized theory defined in the region $r \geq r_{\rm s}(t)$.
For a sketch of the trajectory of the renormalized leaves, see Fig.~\ref{fig:flow}.

We expect that the exterior description presented here can be generalized to the case of arbitrary (not necessarily spherically symmetric) collapsing matter.
The basic idea is to stretch the event horizon in such a way that the system of the stretched and nonstretched horizons appears locally Rindler.
One possible procedure to extend the stretch horizon to earlier times is as follows:\ consider a congruence of past-directed timelike curves $x_p^\mu(\tau)$, each of which starts from a point $p$ on the stretched horizon at a late time $t_0$ in the direction orthogonal to the equal-time hypersurface $t = t_0$.
Here, $\tau$ is the proper time along each curve, with $\tau = 0$ on the stretched horizon.
We require that each curve maintains proper acceleration $a^\mu$ of
\begin{equation}
  \sqrt{a^\mu a_\mu} \approx \frac{1}{l_{\rm s}}
\end{equation}
in the direction orthogonal to the hypersurface given by the spacetime trajectory of the congruence.
An equal-time hypersurface in the exterior spacetime region can be chosen in such a way that it agrees with a constant $\tau$ surface on the congruence.
If caustics develop in the congruence, we terminate the corresponding curves there.
We expect that a procedure like this works, at least, for simple spacetimes, but we leave a detailed study of the stretching of the event horizon for future work.

\subsection{Significance of the event horizon}
\label{subsec:significance}

The discerning reader may note that the argument presented in Section~\ref{subsec:trivial-causal} relies primarily on the holographic principle, and may be applied to any timelike surface outside the event horizon, not just the stretched event horizon.
For example, one might consider in the setup of Fig.~\ref{fig:formation} an extension of the stretched horizon defined by some time-dependent radius $r_{\rm t}(t)$, whose spacetime trajectory is timelike and which agrees with the black hole horizon at the infalling null shell, such as
\begin{equation}
  \frac{d r_{\rm t}(t)}{dt} > 0,
\qquad
  r_{\rm t}(t_{\rm null}) = r_{\rm s}.
\end{equation}
Here, $t_{\rm null}$ is the time at which the trajectory $r_{\rm t}(t)$ of the surface intersects the null shell, and $r_{\rm s}$ is the radius of the stretched horizon of the black hole of mass $M$.
We call this trajectory the ``tube.''
One can imagine that the tube is capped off in the past at some time $t_0$:\ $r_{\rm t}(t_0) = 0$.

Such a description may indeed exist, but the dynamics on the tube would then be very different from the dynamics of the horizon.
To see this, consider a trajectory of a particle which enters and then leaves the tube in the global spacetime picture.
In the exterior picture, this should be described as a process in which a particle hits the surface of the tube at some time and then some time later the same particle is spit out from a different point on the surface.
This implies that, unlike the case of the stretched horizon, the dynamics of the surface are not scrambling and chaotic.
We do, however, expect that these properties are associated with the (stretched) null surface, since a particle inside the surface becomes causally disconnected from the exterior.

What is the significance, then, of the event horizon compared with other null surfaces?
For these other surfaces, such as the horizons of the standard and spherical Rindler spaces, the area of the horizon is either infinite or increases without bound.
We thus expect that information absorbed into the corresponding stretched horizon will not be sent back to the ambient space, since the horizon degrees of freedom will not be fully scrambled.
On the other hand, for an event horizon, its stretched surface becomes the stretched horizon of the black hole, whose area will eventually shrink to zero due to Hawking radiation.
Any information contained within the stretched event horizon, therefore, will be transferred to ambient space at late times, either in the form of Hawking radiation or through mining.

The expectation given above is consistent with the fact that neither standard nor spherical Rindler space contains a singularity.
In these spacetimes, an ``exterior'' observer cannot retrieve information about an object which crossed the horizon at an earlier time.
Thus, there is no problem of cloning even if the observer jumps into the horizon and catches up with the original object, which is generally possible if there is no singularity.
On the other hand, for an event horizon, an exterior observer at late times \textit{can} retrieve information behind the horizon.
Therefore, an observer must not be able to jump into the horizon and obtain a second copy of the information they have retrieved.
This requires that every causal curve in the interior region ends at a singularity, as is indeed the case for a Schwarzschild black hole (and presumably for other unstable black holes as well, after appropriate quantum effects are taken into account).

\subsection{Topology change and quantum global hyperbolicity}
\label{subsec:quant-hyperb}

In the exterior description given here, the flow of information is causally trivial.
This description, however, does not include the region inside the black hole.
This requires that the spatial topology changes as time progresses.
For example, if the black hole is formed in $(d+1)$-dimensional asymptotically flat spacetime, the spatial topologies in the periods before the formation of the stretched event horizon, during the presence of the stretched (event and black hole) horizon, and after the black hole evaporation are $\mathbb{R}^d$, $\mathbb{R}^d \,\backslash\, B^d$, and $\mathbb{R}^d$, respectively.
These spacetimes are not globally hyperbolic classically.
This is due to the singularity at the endpoint of black hole evaporation, and the fact that equal-time hypersurfaces do not enter into the interior of the stretched horizon.

The unitarity of time evolution in the exterior description, however, implies that the spacetime is ``globally hyperbolic'' at the fully quantum level, i.e.\ if we include intrinsically quantum gravitational degrees of freedom on the horizon and use the microscopic Hamiltonian for time evolution.
In this sense, equal-time hypersurfaces in the exterior description play the role of Cauchy surfaces in quantum gravity, and may be called quantum Cauchy surfaces.

\section{Global Spacetime as a Coarse-Grained Description}
\label{sec:global}

We have argued that the global spacetime is only emergent at the level of semiclassical gravity and cannot be used to formulate unitary quantum mechanics at the fundamental level.%
\footnote{
 In the context of AdS/CFT, this implies that any putative nonperturbative bulk theory should not be formulated on a global slice (or should not include a smooth interior), unless an intricate nonlocal mechanism reducing the redundancies on an equal-time hypersurface is involved.
}
On the other hand, the concept of global spacetime itself is well defined in the sense that the physics occurring there is robust under physical processes, including nonperturbative information extraction processes performed on Hawking radiation.
In this section, we discuss how the description based on the global description can emerge as a coarse-grained description, and explain why global methods nevertheless yield important results, including hints of unitarity.

\begin{figure}[t]
\centering
  \includegraphics[height=0.37\textheight]{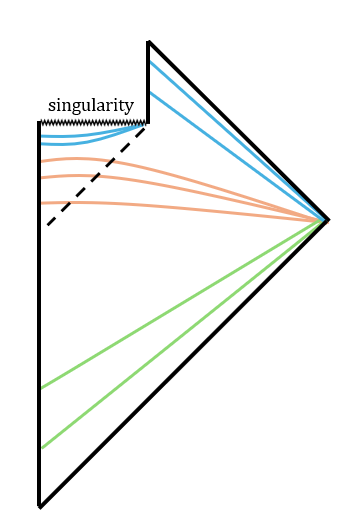}
\vspace{0mm}
\caption{
 The global spacetime description can be obtained by evolving the initial equal-time hypersurface in such a way that at each time, the equal-time hypersurface is a Cauchy slice.
 This description, however, is consistent only if we disallow a process involving an exponentially fine-tuned detector configuration accessing information about the interior; see Section~\ref{sec:puzzle-sol}.
}
\label{fig:global-slice}
\end{figure}

The global spacetime description can be obtained at the semiclassical level by evolving the initial equal-time hypersurface in such a way that at each time, the equal-time hypersurface is a Cauchy slice; see Fig.~\ref{fig:global-slice}.%
\footnote{
 Strictly speaking, equal-time hypersurfaces after black hole evaporation are not Cauchy surfaces at the classical level because of the singularity at the endpoint of the evaporation; see the discussion in Section~\ref{subsec:quant-hyperb}.
}
The famous calculation of Hawking~\cite{Hawking:1976ra} suggests that this corresponds to one which is coarse-grained over black hole microstates.
This results in the final state of Hawking radiation being thermal.
In other words, the global spacetime description regards states after the black hole formation as the (maximal) mixture over black hole microstates~\cite{Langhoff:2020jqa}.
We will show that this interpretation is consistent with results obtained using replica wormholes~\cite{Penington:2019kki,Almheiri:2019qdq}, entanglement islands~\cite{Penington:2019npb,Almheiri:2019psf,Almheiri:2019hni}, and baby universes~\cite{Marolf:2020rpm,Polchinski:1994zs}, which see some traits of unitarity despite the fact that they employ the global spacetime description.

\subsection{The Page curve from replica wormholes and entanglement islands}
\label{subsec:replica-wh}

While the global spacetime description represents an ensemble average over microstates, it can still allow us to calculate the ensemble averages of many different quantities by changing the boundary conditions.%
\footnote{
 In lower dimensional gravity, the relevant ensemble can consist of different microscopic boundary theories~\cite{Saad:2019lba,Stanford:2019vob}.
}
In particular, by adopting the replica boundary conditions~\cite{Penington:2019kki,Almheiri:2019qdq}, gravitational path integrals in the global spacetime can calculate the ensemble averages of the traces of powers of the density matrix of Hawking radiation $\overline{\Tr \rho_R^n}$ for different $n$.
From this, the ensemble average of the von~Neumann entropy of the radiation can be obtained as
\begin{equation}
  \overline{S_R} = - \lim_{n \rightarrow 1} \frac{\partial}{\partial n} \overline{\Tr \rho_R^n} \sim {\rm min}\{ S_{\rm rad}, S_{\rm bh} \},
\end{equation}
where $S_{\rm rad}$ and $S_{\rm bh}$ are the coarse-grained entropies of the radiation and the black hole.
Since this quantity is the ensemble average of microscopic von~Neumann entropies, which obey the Page curve~\cite{Page:1993wv} for all members of the ensemble, it also obeys the Page curve~\cite{Langhoff:2020jqa,Bousso:2020kmy,Engelhardt:2020qpv,Renner:2021qbe,Qi:2021oni,Almheiri:2021jwq,Blommaert:2021fob,Chandra:2022fwi}.
This is how the replica wormhole calculation in~\cite{Penington:2019kki,Almheiri:2019qdq} yields the Page curve, despite the fact that it employs the global spacetime description, which is not unitary.%
\footnote{
 To recover the full unitary content of the theory, one would calculate the ensemble averages of all of the moments, $\overline{\Tr \rho_R^n}$ for $1 \leq n \leq e^{S_{\rm rad}}$.
 Such a calculation, however, would not be able to be performed within the framework of semiclassical gravity.
}

The replica prescription for the gravitational path integral is virtually equivalent to the entanglement island prescription~\cite{Penington:2019npb,Almheiri:2019psf,Almheiri:2019hni} for calculating entropies.
Thus, the success of the calculation in~\cite{Penington:2019npb,Almheiri:2019psf,Almheiri:2019hni}, which adopts the global spacetime description, can also be interpreted in the same way.
Note that because of the ensemble averaging, Hawking radiation in a single copy of spacetime does not contain information about the microstate by itself.
This is why the semiclassical entropy must be used for Hawking radiation when applying the island formula (in a single copy of the global spacetime).

\subsection{Baby universe \texorpdfstring{\boldmath $\alpha$}{TEXT}-states as fine-grained information in Hawking radiation}
\label{subsec:baby-u}

\begin{figure}[t]
\centering
  \includegraphics[height=0.32\textheight]{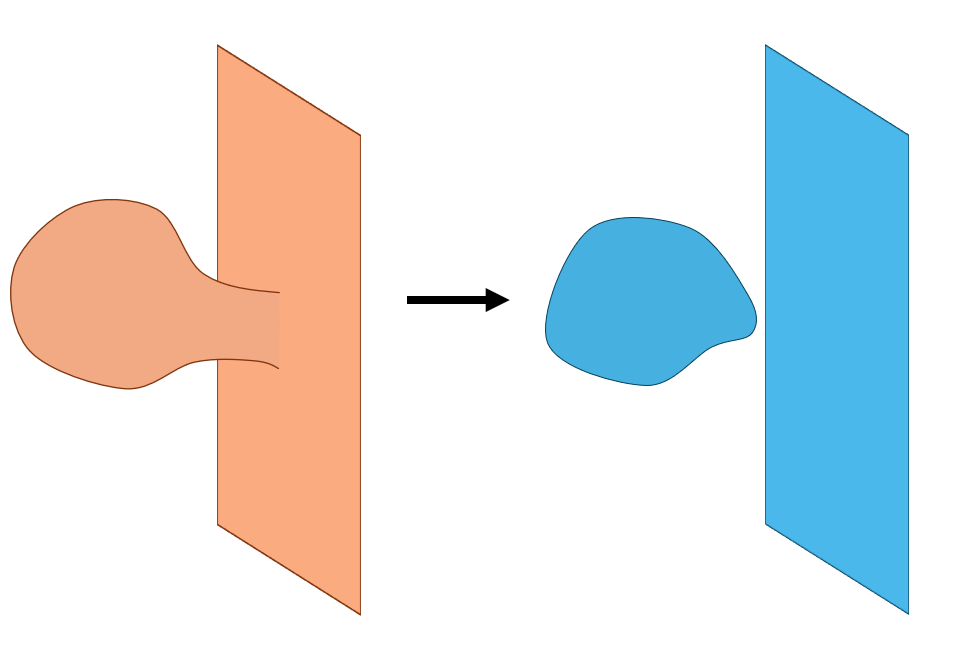}
\vspace{-5mm}
\caption{
 Assuming that the singularity at the endpoint of black hole evaporation is sufficiently weak, the evaporation of a black hole can be viewed as the creation of a baby universe in (but only in) the effective global spacetime description.
}
\label{fig:baby-u}
\end{figure}

The transition from connected to disconnected Cauchy surfaces depicted in Fig.~\ref{fig:global-slice} can be viewed as the creation of a baby universe~\cite{Coleman:1988cy,Giddings:1988cx,Giddings:1988wv,Polchinski:1994zs}, which will be separate from the ambient space at late times.
This picture was studied by Marolf and Maxfield in~\cite{Marolf:2020rpm}.
Assuming that the singularity at the end of the black hole evaporation is mild, we may envision the process as in Fig.~\ref{fig:baby-u}, in which a closed universe corresponding to the black hole interior is separated from the ambient space describing the exterior of the black hole.

Marolf's and Maxfield's interpretation is that after the black hole evaporates, its microscopic information is encoded in the $\alpha$-states of the baby universes arising in the system of replicated black holes.
For a single black hole, different $\alpha$ sectors comprise superselection sectors, and hence its information cannot be accessed in a single experiment involving a single black hole.
Our analysis implies that this is an artifact of the global spacetime description, in which the Hawking radiation involves the ensemble average over the microstates.
In the full exterior description, the microscopic information \textit{can} be accessed even in a single experiment by an appropriately fine-tuned detector, as depicted in Fig.~\ref{fig:clone-slice}.
In other words, $\alpha$ sectors in the global spacetime description are not superselection sectors in the exterior description, which is unitary and hence, unlike~\cite{Marolf:2020rpm}, gives a pure final state.

This interpretation is consistent with the conjecture~\cite{McNamara:2020uza,Marolf:2020xie} that in a complete theory of quantum gravity, the Hilbert space of baby universes is trivial
\begin{equation}
  {\rm dim} {\cal H}_{\rm BU} = 1.
\end{equation}
A nontrivial Hilbert space for baby universe states emerges only in an effective description in which certain microscopic information in the fundamental description is coarse-grained.

We expect that this conclusion applies more generally.
For example, we may imagine that states representing a semiclassical closed universe can be embedded, possibly like bound states~\cite{Nomura:2012zb,Maltz:2016iaw}, in a theory accommodating an asymptotic region (such as null infinities of flat space and the conformal boundary of AdS space).
Different $\alpha$ sectors associated with the semiclassical closed universe would then correspond to different microstates in the putative boundary theory in which this universe is embedded.

\section{Conclusion}
\label{sec:concl}

In this paper we have described a consistent view of black hole complementarity.
In this picture, the fundamental description of the spacetime ends at the stretched horizon.
A semiclassical picture of spacetime including the interior can be obtained only effectively, by the mechanism described in Section~\ref{sec:emergence}.
As discussed there, the interior emerges as a way of reorganizing the degrees of freedom participating in the transient phenomenon of an infalling object dissipating into the stretched horizon.
We argued using this framework that no operation performed on Hawking radiation, however complex, can affect the experience of an infalling observer.
Rather, reconstructions of interior operators from Hawking radiation should be interpreted as acting on a historical record of the effective interior, which has causally propagated from the vicinity of the stretched horizon (and early Hawking radiation).
While in the effective global picture the interior operators appear spacelike separated from the exterior operators used for reconstruction, in the fundamental picture they are timelike related.
We explained this phenomenon as a consequence of the chaotic dynamics of the horizon modes.

It is natural to ask whether this construction can be extended to describe a black hole formed from collapse.
To do so, it would be necessary to extend the stretched horizon before the formation of the black hole.
In Section~\ref{sec:causal}, we discussed a scheme for doing this, in which the extended stretched horizon is defined as a fixed radius surface in spherical Rindler coordinates, where the Unruh temperature reaches the string scale.
We speculate, by analogy to the black hole case, that degrees of freedom in the interior might be constructed using the Unruh modes near this extended stretched horizon.
We leave exploration of this and extension to spacetimes without spherical symmetry for future work.

\section*{Acknowledgments}

We thank Masahiro Hotta and Tomonori Ugajin for useful discussions.
This work was supported in part by the Department of Energy, Office of Science, Office of High Energy Physics under QuantISED award DE-SC0019380 and contract DE-AC02-05CH11231 and in part by MEXT KAKENHI grant number JP20H05850, JP20H05860.


\begin{thebibliography}{99}

\bibitem{Bekenstein:1973ur}
J.~D.~Bekenstein,
``Black holes and entropy,''
Phys. Rev. D \textbf{7}, 2333 (1973).

\bibitem{Bekenstein:1974ax}
J.~D.~Bekenstein,
``Generalized second law of thermodynamics in black hole physics,''
Phys. Rev. D \textbf{9}, 3292 (1974).

\bibitem{Hawking:1974sw}
S.~W.~Hawking,
``Particle creation by black holes,''
Commun. Math. Phys. \textbf{43}, 199 (1975)
[Erratum:\ Commun.\ Math.\ Phys.\ {\bf 46}, 206 (1976)].

\bibitem{Susskind:1993if}
L.~Susskind, L.~Thorlacius and J.~Uglum,
``The stretched horizon and black hole complementarity,''
Phys. Rev. D \textbf{48}, 3743 (1993)
[arXiv:hep-th/9306069].

\bibitem{Penington:2019npb}
G.~Penington,
``Entanglement wedge reconstruction and the information paradox,''
JHEP \textbf{09}, 002 (2020)
[arXiv:1905.08255 [hep-th]].

\bibitem{Almheiri:2019psf}
A.~Almheiri, N.~Engelhardt, D.~Marolf and H.~Maxfield,
``The entropy of bulk quantum fields and the entanglement wedge of an evaporating black hole,''
JHEP \textbf{12}, 063 (2019)
[arXiv:1905.08762 [hep-th]].

\bibitem{Almheiri:2019hni}
A.~Almheiri, R.~Mahajan, J.~Maldacena and Y.~Zhao,
``The Page curve of Hawking radiation from semiclassical geometry,''
JHEP \textbf{03}, 149 (2020)
[arXiv:1908.10996 [hep-th]].

\bibitem{Penington:2019kki}
G.~Penington, S.~H.~Shenker, D.~Stanford and Z.~Yang,
``Replica wormholes and the black hole interior,''
JHEP \textbf{03}, 205 (2022)
[arXiv:1911.11977 [hep-th]].

\bibitem{Almheiri:2019qdq}
A.~Almheiri, T.~Hartman, J.~Maldacena, E.~Shaghoulian and A.~Tajdini,
``Replica wormholes and the entropy of Hawking radiation,''
JHEP \textbf{05}, 013 (2020)
[arXiv:1911.12333 [hep-th]].

\bibitem{Ryu:2006bv}
S.~Ryu and T.~Takayanagi,
``Holographic derivation of entanglement entropy from AdS/CFT,''
Phys. Rev. Lett. \textbf{96}, 181602 (2006)
[arXiv:hep-th/0603001].

\bibitem{Hubeny:2007xt}
V.~E.~Hubeny, M.~Rangamani and T.~Takayanagi,
``A covariant holographic entanglement entropy proposal,''
JHEP \textbf{07}, 062 (2007)
[arXiv:0705.0016 [hep-th]].

\bibitem{Faulkner:2013ana}
T.~Faulkner, A.~Lewkowycz and J.~Maldacena,
``Quantum corrections to holographic entanglement entropy,''
JHEP \textbf{11}, 074 (2013)
[arXiv:1307.2892 [hep-th]].

\bibitem{Engelhardt:2014gca}
N.~Engelhardt and A.~C.~Wall,
``Quantum extremal surfaces:\ holographic entanglement entropy beyond the classical regime,''
JHEP \textbf{01}, 073 (2015)
[arXiv:1408.3203 [hep-th]].

\bibitem{tHooft:1993dmi}
G.~'t Hooft,
``Dimensional reduction in quantum gravity,''
in {\it Salamfestschrift},
edited by A.~Ali, J.~Ellis, and S.~Randjbar-Daemi
(World Scientific, Singapore, 1994), p.~284
[arXiv:gr-qc/9310026].

\bibitem{Susskind:1994vu}
L.~Susskind,
``The world as a hologram,''
J. Math. Phys. \textbf{36}, 6377 (1995)
[arXiv:hep-th/9409089].

\bibitem{Maldacena:1997re}
J.~M.~Maldacena,
``The large N limit of superconformal field theories and supergravity,''
Int. J. Theor. Phys. \textbf{38}, 1113 (1999)
[arXiv:hep-th/9711200].

\bibitem{Czech:2012bh}
B.~Czech, J.~L.~Karczmarek, F.~Nogueira and M.~Van Raamsdonk,
``The gravity dual of a density matrix,''
Class. Quant. Grav. \textbf{29}, 155009 (2012)
[arXiv:1204.1330 [hep-th]].

\bibitem{Wall:2012uf}
A.~C.~Wall,
``Maximin surfaces, and the strong subadditivity of the covariant holographic entanglement entropy,''
Class. Quant. Grav. \textbf{31}, 225007 (2014)
[arXiv:1211.3494 [hep-th]].

\bibitem{Headrick:2014cta}
M.~Headrick, V.~E.~Hubeny, A.~Lawrence and M.~Rangamani,
``Causality \& holographic entanglement entropy,''
JHEP \textbf{12}, 162 (2014)
[arXiv:1408.6300 [hep-th]].

\bibitem{Jafferis:2015del}
D.~L.~Jafferis, A.~Lewkowycz, J.~Maldacena and S.~J.~Suh,
``Relative entropy equals bulk relative entropy,''
JHEP \textbf{06}, 004 (2016)
[arXiv:1512.06431 [hep-th]].

\bibitem{Dong:2016eik}
X.~Dong, D.~Harlow and A.~C.~Wall,
``Reconstruction of bulk operators within the entanglement wedge in gauge-gravity duality,''
Phys. Rev. Lett. \textbf{117}, 021601 (2016)
[arXiv:1601.05416 [hep-th]].

\bibitem{Faulkner:2017vdd}
T.~Faulkner and A.~Lewkowycz,
``Bulk locality from modular flow,''
JHEP \textbf{07}, 151 (2017)
[arXiv:1704.05464 [hep-th]].

\bibitem{Cotler:2017erl}
J.~Cotler, P.~Hayden, G.~Penington, G.~Salton, B.~Swingle and M.~Walter,
``Entanglement wedge reconstruction via universal recovery channels,''
Phys. Rev. X \textbf{9}, 031011 (2019)
[arXiv:1704.05839 [hep-th]].

\bibitem{Akers:2020pmf}
C.~Akers and G.~Penington,
``Leading order corrections to the quantum extremal surface prescription,''
JHEP \textbf{04}, 062 (2021)
[arXiv:2008.03319 [hep-th]].

\bibitem{Kim:2020cds}
I.~Kim, E.~Tang and J.~Preskill,
``The ghost in the radiation:\ robust encodings of the black hole interior,''
JHEP \textbf{06}, 031 (2020)
[arXiv:2003.05451 [hep-th]].

\bibitem{Bousso:2023sya}
R.~Bousso and G.~Penington,
``Holograms in our world,''
Phys. Rev. D \textbf{108}, 046007 (2023)
[arXiv:2302.07892 [hep-th]].

\bibitem{Nomura:2018kia}
Y.~Nomura,
``Reanalyzing an evaporating black hole,''
Phys. Rev. D \textbf{99}, 086004 (2019)
[arXiv:1810.09453 [hep-th]].

\bibitem{Nomura:2019qps}
Y.~Nomura,
``Spacetime and universal soft modes --- black holes and beyond,''
Phys. Rev. D \textbf{101}, 066024 (2020)
[arXiv:1908.05728 [hep-th]].

\bibitem{Nomura:2019dlz}
Y.~Nomura,
``The interior of a unitarily evaporating black hole,''
Phys. Rev. D \textbf{102}, 026001 (2020)
[arXiv:1911.13120 [hep-th]].

\bibitem{Langhoff:2020jqa}
K.~Langhoff and Y.~Nomura,
``Ensemble from coarse graining:\ reconstructing the interior of an evaporating black hole,''
Phys. Rev. D \textbf{102}, 086021 (2020)
[arXiv:2008.04202 [hep-th]].

\bibitem{Nomura:2020ska}
Y.~Nomura,
``Black hole interior in unitary gauge construction,''
Phys. Rev. D \textbf{103}, 066011 (2021)
[arXiv:2010.15827 [hep-th]].

\bibitem{Murdia:2022giv}
C.~Murdia, Y.~Nomura and K.~Ritchie,
``Black hole and de~Sitter microstructures from a semiclassical perspective,''
Phys. Rev. D \textbf{107}, 026016 (2023)
[arXiv:2207.01625 [hep-th]].

\bibitem{Unruh:1982ic}
W.~G.~Unruh and R.~M.~Wald,
``Acceleration radiation and generalized second law of thermodynamics,''
Phys. Rev. D \textbf{25}, 942 (1982).

\bibitem{Brown:2012un}
A.~R.~Brown,
``Tensile strength and the mining of black holes,''
Phys. Rev. Lett. \textbf{111}, 211301 (2013)
[arXiv:1207.3342 [gr-qc]].

\bibitem{Wootters:1982zz}
W.~K.~Wootters and W.~H.~Zurek,
``A single quantum cannot be cloned,''
Nature \textbf{299}, 802 (1982).

\bibitem{Coleman:1988cy}
S.~Coleman,
``Black holes as red herrings:\ topological fluctuations and the loss of quantum coherence,''
Nucl. Phys. B \textbf{307}, 867 (1988).

\bibitem{Giddings:1988cx}
S.~B.~Giddings and A.~Strominger,
``Loss of incoherence and determination of coupling constants in quantum gravity,''
Nucl. Phys. B \textbf{307}, 854 (1988).

\bibitem{Giddings:1988wv}
S.~B.~Giddings and A.~Strominger,
``Baby universes, third quantization and the cosmological constant,''
Nucl. Phys. B \textbf{321}, 481 (1989).

\bibitem{Marolf:2020rpm}
D.~Marolf and H.~Maxfield,
``Observations of Hawking radiation:\ the Page curve and baby universes,''
JHEP \textbf{04}, 272 (2021)
[arXiv:2010.06602 [hep-th]].

\bibitem{Polchinski:1994zs}
J.~Polchinski and A.~Strominger,
``A possible resolution of the black hole information puzzle,''
Phys. Rev. D \textbf{50}, 7403 (1994)
[arXiv:hep-th/9407008].

\bibitem{tHooft:1984kcu}
G.~'t Hooft,
``On the quantum structure of a black hole,''
Nucl. Phys. B \textbf{256}, 727 (1985).

\bibitem{tHooft:1990fkf}
G.~'t Hooft,
``The black hole interpretation of string theory,''
Nucl. Phys. B \textbf{335}, 138 (1990).

\bibitem{Page:1993wv}
D.~N.~Page,
``Information in black hole radiation,''
Phys. Rev. Lett. \textbf{71}, 3743 (1993)
[arXiv:hep-th/9306083].

\bibitem{Nomura:2018kji}
Y.~Nomura, P.~Rath and N.~Salzetta,
``Pulling the boundary into the bulk,''
Phys. Rev. D \textbf{98}, 026010 (2018)
[arXiv:1805.00523 [hep-th]].

\bibitem{Murdia:2020iac}
C.~Murdia, Y.~Nomura and P.~Rath,
``Coarse-graining holographic states:\ a semiclassical flow in general spacetimes,''
Phys. Rev. D \textbf{102}, 086001 (2020)
[arXiv:2008.01755 [hep-th]].

\bibitem{Lunin:2001jy}
O.~Lunin and S.~D.~Mathur,
``AdS/CFT duality and the black hole information paradox,''
Nucl. Phys. B \textbf{623}, 342 (2002)
[arXiv:hep-th/0109154].

\bibitem{Lunin:2002iz}
O.~Lunin, J.~M.~Maldacena and L.~Maoz,
``Gravity solutions for the D1-D5 system with angular momentum,''
arXiv:hep-th/0212210.

\bibitem{Mathur:2003hj}
S.~D.~Mathur, A.~Saxena and Y.~K.~Srivastava,
``Constructing `hair' for the three charge hole,''
Nucl. Phys. B \textbf{680}, 415 (2004)
[arXiv:hep-th/0311092].

\bibitem{Almheiri:2012rt}
A.~Almheiri, D.~Marolf, J.~Polchinski and J.~Sully,
``Black holes:\ complementarity or firewalls?,''
JHEP \textbf{02}, 062 (2013)
[arXiv:1207.3123 [hep-th]].

\bibitem{Almheiri:2013hfa}
A.~Almheiri, D.~Marolf, J.~Polchinski, D.~Stanford and J.~Sully,
``An apologia for firewalls,''
JHEP \textbf{09}, 018 (2013)
[arXiv:1304.6483 [hep-th]].

\bibitem{Marolf:2013dba}
D.~Marolf and J.~Polchinski,
``Gauge/gravity duality and the black hole interior,''
Phys. Rev. Lett. \textbf{111}, 171301 (2013)
[arXiv:1307.4706 [hep-th]].

\bibitem{Hayden:2007cs}
P.~Hayden and J.~Preskill,
``Black holes as mirrors:\ quantum information in random subsystems,''
JHEP \textbf{09}, 120 (2007)
[arXiv:0708.4025 [hep-th]].

\bibitem{Sekino:2008he}
Y.~Sekino and L.~Susskind,
``Fast scramblers,''
JHEP \textbf{10}, 065 (2008)
[arXiv:0808.2096 [hep-th]].

\bibitem{Maldacena:2015waa}
J.~Maldacena, S.~H.~Shenker and D.~Stanford,
``A bound on chaos,''
JHEP \textbf{08}, 106 (2016)
[arXiv:1503.01409 [hep-th]].

\bibitem{Papadodimas:2012aq}
K.~Papadodimas and S.~Raju,
``An infalling observer in AdS/CFT,''
JHEP \textbf{10}, 212 (2013)
[arXiv:1211.6767 [hep-th]].

\bibitem{Verlinde:2012cy}
E.~Verlinde and H.~Verlinde,
``Black hole entanglement and quantum error correction,''
JHEP \textbf{10}, 107 (2013)
[arXiv:1211.6913 [hep-th]].

\bibitem{Nomura:2012ex}
Y.~Nomura and J.~Varela,
``A note on (no) firewalls:\ the entropy argument,''
JHEP \textbf{07}, 124 (2013)
[arXiv:1211.7033 [hep-th]].

\bibitem{Maldacena:2013xja}
J.~Maldacena and L.~Susskind,
``Cool horizons for entangled black holes,''
Fortsch. Phys. \textbf{61}, 781 (2013)
[arXiv:1306.0533 [hep-th]].

\bibitem{Papadodimas:2013jku}
K.~Papadodimas and S.~Raju,
``State-dependent bulk-boundary maps and black hole complementarity,''
Phys. Rev. D \textbf{89}, 086010 (2014)
[arXiv:1310.6335 [hep-th]].

\bibitem{Papadodimas:2015jra}
K.~Papadodimas and S.~Raju,
``Remarks on the necessity and implications of state-dependence in the black hole interior,''
Phys. Rev. D \textbf{93}, 084049 (2016)
[arXiv:1503.08825 [hep-th]].

\bibitem{Hayden:2018khn}
P.~Hayden and G.~Penington,
``Learning the alpha-bits of black holes,''
JHEP \textbf{12}, 007 (2019)
[arXiv:1807.06041 [hep-th]].

\bibitem{Mathur:2009hf}
S.~D.~Mathur,
``The information paradox:\ a pedagogical introduction,''
Class. Quant. Grav. \textbf{26}, 224001 (2009)
[arXiv:0909.1038 [hep-th]].

\bibitem{Bousso:2013ifa}
R.~Bousso,
``Violations of the equivalence principle by a nonlocally reconstructed vacuum at the black hole horizon,''
Phys. Rev. Lett. \textbf{112}, 041102 (2014)
[arXiv:1308.3697 [hep-th]].

\bibitem{Almheiri:2014lwa}
A.~Almheiri, X.~Dong and D.~Harlow,
``Bulk locality and quantum error correction in AdS/CFT,''
JHEP \textbf{04}, 163 (2015)
[arXiv:1411.7041 [hep-th]].

\bibitem{Harlow:2013tf}
D.~Harlow and P.~Hayden,
``Quantum computation vs.\ firewalls,''
JHEP \textbf{06}, 085 (2013)
[arXiv:1301.4504 [hep-th]].

\bibitem{Brown:2019rox}
A.~R.~Brown, H.~Gharibyan, G.~Penington and L.~Susskind,
``The python\textquoteright{}s lunch: geometric obstructions to decoding Hawking radiation,''
JHEP \textbf{08}, 121 (2020)
[arXiv:1912.00228 [hep-th]].

\bibitem{Bousso:2022hlz}
R.~Bousso and G.~Penington,
``Entanglement wedges for gravitating regions,''
Phys. Rev. D \textbf{107}, 086002 (2023)
[arXiv:2208.04993 [hep-th]].

\bibitem{Lowe:1995ac}
D.~A.~Lowe, J.~Polchinski, L.~Susskind, L.~Thorlacius and J.~Uglum,
``Black hole complementarity versus locality,''
Phys. Rev. D \textbf{52}, 6997 (1995)
[arXiv:hep-th/9506138].

\bibitem{Giddings:2006sj}
S.~B.~Giddings,
``Black hole information, unitarity, and nonlocality,''
Phys. Rev. D \textbf{74}, 106005 (2006)
[arXiv:hep-th/0605196].

\bibitem{Bousso:2022tdb}
R.~Bousso and A.~Shahbazi-Moghaddam,
``Quantum singularities,''
Phys. Rev. D \textbf{107}, 066002 (2023)
[arXiv:2206.07001 [hep-th]].

\bibitem{Nomura:2011rb}
Y.~Nomura,
``Quantum mechanics, spacetime locality, and gravity,''
Found. Phys. \textbf{43}, 978 (2013)
[arXiv:1110.4630 [hep-th]].

\bibitem{q-Darwinism}
H.~Ollivier, D.~Poulin and W.~H.~Zurek,
``Objective properties from subjective quantum states:\ environment as a witness,''
Phys.\ Rev.\ Lett.\ {\bf 93}, 220401 (2004)
[arXiv:quant-ph/0307229].

\bibitem{q-Darwinism-2}
R.~Blume-Kohout and W.~H.~Zurek,
``Quantum Darwinism:\ entanglement, branches, and the emergent classicality of redundantly stored quantum information,''
Phys.\ Rev.\ A {\bf 73}, 062310 (2006)
[arXiv:quant-ph/0505031].

\bibitem{Bousso:2011up}
R.~Bousso and L.~Susskind,
``The multiverse interpretation of quantum mechanics,''
Phys. Rev. D \textbf{85}, 045007 (2012)
[arXiv:1105.3796 [hep-th]].

\bibitem{Zurek:1982zz}
W.~H.~Zurek,
``Entropy evaporated by a black hole,''
Phys. Rev. Lett. \textbf{49}, 1683 (1982).

\bibitem{Page:1983ug}
D.~N.~Page,
``Comment on `Entropy evaporated by a black hole',''
Phys. Rev. Lett. \textbf{50}, 1013 (1983).

\bibitem{Page:1979tc}
D.~N.~Page,
``Is black hole evaporation predictable?,''
Phys. Rev. Lett. \textbf{44}, 301 (1980).

\bibitem{Nomura:2012cx}
Y.~Nomura, J.~Varela and S.~J.~Weinberg,
``Black holes, information, and Hilbert space for quantum gravity,''
Phys. Rev. D \textbf{87}, 084050 (2013)
[arXiv:1210.6348 [hep-th]].

\bibitem{Bouland:2019pvu}
A.~Bouland, B.~Fefferman and U.~Vazirani,
``Computational pseudorandomness, the wormhole growth paradox, and constraints on the AdS/CFT duality,''
arXiv:1910.14646 [quant-ph].

\bibitem{Engelhardt:2024hpe}
N.~Engelhardt, \r{A}.~Folkestad, A.~Levine, E.~Verheijden and L.~Yang,
``Cryptographic censorship,''
arXiv:2402.03425 [hep-th].

\bibitem{Bousso:2002ju}
R.~Bousso,
``The holographic principle,''
Rev. Mod. Phys. \textbf{74}, 825 (2002)
[arXiv:hep-th/0203101].

\bibitem{Balasubramanian:2013rqa}
V.~Balasubramanian, B.~Czech, B.~D.~Chowdhury and J.~de Boer,
``The entropy of a hole in spacetime,''
JHEP \textbf{10}, 220 (2013)
[arXiv:1305.0856 [hep-th]].

\bibitem{Unruh:1976db}
W.~G.~Unruh,
``Notes on black hole evaporation,''
Phys. Rev. D \textbf{14}, 870 (1976).

\bibitem{Fulling:1972md}
S.~A.~Fulling,
``Nonuniqueness of canonical field quantization in Riemannian space-time,''
Phys.\ Rev.\ D {\bf 7}, 2850 (1973).

\bibitem{Davies:1974th}
P.~C.~W.~Davies,
``Scalar particle production in Schwarzschild and Rindler metrics,''
J.\ Phys.\ A {\bf 8}, 609 (1975).

\bibitem{Parikh:2017aas}
M.~Parikh and A.~Svesko,
``Einstein's equations from the stretched future light cone,''
Phys. Rev. D \textbf{98}, 026018 (2018)
[arXiv:1712.08475 [hep-th]].

\bibitem{Swingle:2009bg}
B.~Swingle,
``Entanglement renormalization and holography,''
Phys. Rev. D \textbf{86}, 065007 (2012)
[arXiv:0905.1317 [cond-mat.str-el]].

\bibitem{Pastawski:2015qua}
F.~Pastawski, B.~Yoshida, D.~Harlow and J.~Preskill,
``Holographic quantum error-correcting codes:\ toy models for the bulk/boundary correspondence,''
JHEP \textbf{06}, 149 (2015)
[arXiv:1503.06237 [hep-th]].

\bibitem{Hayden:2016cfa}
P.~Hayden, S.~Nezami, X.-L.~Qi, N.~Thomas, M.~Walter and Z.~Yang,
``Holographic duality from random tensor networks,''
JHEP \textbf{11}, 009 (2016)
[arXiv:1601.01694 [hep-th]].

\bibitem{Miyaji:2015yva}
M.~Miyaji and T.~Takayanagi,
``Surface/state correspondence as a generalized holography,''
PTEP \textbf{2015}, 073B03 (2015)
[arXiv:1503.03542 [hep-th]].

\bibitem{Bousso:2015mna}
R.~Bousso, Z.~Fisher, S.~Leichenauer and A.~C.~Wall,
``Quantum focusing conjecture,''
Phys. Rev. D \textbf{93}, 064044 (2016)
[arXiv:1506.02669 [hep-th]].

\bibitem{Hawking:1976ra}
S.~W.~Hawking,
``Breakdown of predictability in gravitational collapse,''
Phys. Rev. D \textbf{14}, 2460 (1976).

\bibitem{Saad:2019lba}
P.~Saad, S.~H.~Shenker and D.~Stanford,
``JT gravity as a matrix integral,''
arXiv:1903.11115 [hep-th].

\bibitem{Stanford:2019vob}
D.~Stanford and E.~Witten,
``JT gravity and the ensembles of random matrix theory,''
Adv. Theor. Math. Phys. \textbf{24}, 1475 (2020)
[arXiv:1907.03363 [hep-th]].

\bibitem{Bousso:2020kmy}
R.~Bousso and E.~Wildenhain,
``Gravity/ensemble duality,''
Phys. Rev. D \textbf{102}, 066005 (2020)
[arXiv:2006.16289 [hep-th]].

\bibitem{Engelhardt:2020qpv}
N.~Engelhardt, S.~Fischetti and A.~Maloney,
``Free energy from replica wormholes,''
Phys. Rev. D \textbf{103}, 046021 (2021)
[arXiv:2007.07444 [hep-th]].

\bibitem{Renner:2021qbe}
R.~Renner and J.~Wang,
``The black hole information puzzle and the quantum de~Finetti theorem,''
arXiv:2110.14653 [hep-th].

\bibitem{Qi:2021oni}
X.-L.~Qi, Z.~Shangnan and Z.~Yang,
``Holevo information and ensemble theory of gravity,''
JHEP \textbf{02}, 056 (2022)
[arXiv:2111.05355 [hep-th]].

\bibitem{Almheiri:2021jwq}
A.~Almheiri and H.~W.~Lin,
``The entanglement wedge of unknown couplings,''
JHEP \textbf{08}, 062 (2022)
[arXiv:2111.06298 [hep-th]].

\bibitem{Blommaert:2021fob}
A.~Blommaert, L.~V.~Iliesiu and J.~Kruthoff,
``Gravity factorized,''
JHEP \textbf{09}, 080 (2022)
[arXiv:2111.07863 [hep-th]].

\bibitem{Chandra:2022fwi}
J.~Chandra and T.~Hartman,
``Coarse graining pure states in AdS/CFT,''
JHEP \textbf{10}, 030 (2023)
[arXiv:2206.03414 [hep-th]].

\bibitem{McNamara:2020uza}
J.~McNamara and C.~Vafa,
``Baby universes, holography, and the swampland,''
arXiv:2004.06738 [hep-th].

\bibitem{Marolf:2020xie}
D.~Marolf and H.~Maxfield,
``Transcending the ensemble:\ baby universes, spacetime wormholes, and the order and disorder of black hole information,''
JHEP \textbf{08}, 044 (2020)
[arXiv:2002.08950 [hep-th]].

\bibitem{Nomura:2012zb}
Y.~Nomura,
``The static quantum multiverse,''
Phys. Rev. D \textbf{86}, 083505 (2012)
[arXiv:1205.5550 [hep-th]].

\bibitem{Maltz:2016iaw}
J.~Maltz and L.~Susskind,
``de Sitter space as a resonance,''
Phys. Rev. Lett. \textbf{118}, 101602 (2017)
[arXiv:1611.00360 [hep-th]].

\end{thebibliography}
\end{document}